\documentclass[12pt]{article}
\usepackage{graphicx}
\usepackage{subfigure}
\newcommand{\beq}{\begin{equation}}
\newcommand{\eeq}{\end{equation}}
\newcommand{\bea}{\begin{eqnarray}}
\newcommand{\eea}{\end{eqnarray}}

\newcommand{\overlrarrow}[1]{\vbox{\ialign{##\cr\cr
                  \leftrightarrowfill\crcr\noalign{\kern-1pt\nointerlineskip}
                  $\hfil\displaystyle{#1}\hfil$\crcr}}}

\begin{document}
\begin{titlepage}
\begin{flushleft}
       \hfill                      {\tt hep-th/0909.5522}\\
       \hfill                       FIT HE - 09-02 \\
       \hfill                       KYUSHU-HET 121 \\
\end{flushleft}
\vspace*{3mm}
\begin{center}
{\bf\LARGE Holographic Confining Gauge theory \\
\vspace*{3mm}
and Response to Electric Field}

\vspace*{5mm}
\vspace*{2mm}
\vspace*{5mm}
{\large Kazuo Ghoroku${}^{\dagger}$\footnote[1]{\tt gouroku@dontaku.fit.ac.jp},
Masafumi Ishihara${}^{\ddagger}$\footnote[2]{\tt masafumi@higgs.phys.kyushu-u.ac.jp},
Tomoki Taminato${}^{\ddagger}$\footnote[2]{\tt taminato@higgs.phys.kyushu-u.ac.jp},
%
}\\

{${}^{\dagger}$Fukuoka Institute of Technology, Wajiro, 
Higashi-ku} \\
{
Fukuoka 811-0295, Japan\\}
{
${}^{\ddagger}$Department of Physics, Kyushu University, Hakozaki,
Higashi-ku}\\
{
Fukuoka 812-8581, Japan\\}

\end{center}

\begin{abstract}

We study the response of confining gauge theory to the external electric field
by using holographic 
Yang-Mills theories in the large $N_c$ limit. Although the theories
are in the confinement phase, we find a transition from the insulator
to the conductor phase when the electric field exceeds its critical value. 
Then, the baryon number current is generated in the conductor phase.
At the same time, in this phase, the meson melting is observed
through the quasi-normal modes of meson spectrum. 
Possible ideas are given for the string state corresponding to the melted mesons,
and they lead to the idea that the source of this current
may be identified with the quarks and anti-quarks supplied by the melted mesons.
We also discuss about other possible carriers.
Furthermore, from the analysis of the massless quark, chiral symmetry 
restoration is observed at the insulator-conductor transition point
by studying a confining theory in which the chiral symmetry is broken. 

\end{abstract}
\end{titlepage}

\section{Introduction}

In the context of the holography \cite{juan,bigRev},
the properties of
flavor quarks have been studied by embedding D7 brane(s) as a probe
 in type IIB theory \cite{KK,KMMW,KMMW2,Bab,ES,SS,NPR,GY,CNP}. 
Recently, the research in this direction has been done by introducing
the external electric field of $U(1)_B$ gauge symmetry, where the charge of the current
is the baryon number (see refs.\cite{Karch:2007pd,O'Bannon:2007in}). And 
many electric properties (such as the conductivity) of the system were uncovered. 
At the same time,
the external magnetic field for the finite temperature case has also been studied
by uncovering the phase diagram, which shows the competition of the temperature and 
the magnetic field \cite{Albash:2007bk}.
The case of external magnetic field at zero temperature was studied in refs.\cite{Filev:2007gb,Filev:2007qu}, where a number of interesting phenomena (such as spontaneous chiral symmetry breaking) were readily extracted \cite{Erdmenger:2007bn}.

\vspace{.3cm}
One of the important electric properties observed in deconfining theories is the phase 
transition from the insulator to the conductor. This transition 
is always seen when the external electric field, how small it is, 
is introduced in the theory even if at zero temperature \cite{AFJK2}.
And in the conductor phase, 
two origins of charge carrier are considered. One is set by the time
component of the gauge field, which introduces the chemical potential and the 
number density ($n_b$)
of quarks. Finite $n_b$ can be introduced only in the black-hole embedded D7 
brane \cite{KO}, which is realized only in the high temperature 
deconfining phase.
Another is considered as the pair creation of quark and 
anti- quark from the vacuum due to the external electric field.
In both cases, the carrier of the baryon number
current is reduced to the free quark and anti-quark, which are the 
strings connecting the probe brane and the horizon.
Even if the temperature were zero, the conducting 
solutions are seen when the external electric field, how small it is, 
is introduced in the deconfining theory \cite{AFJK2}. This case is considered
as the limit of finite temperature deconfining phase.

In this paper, instead, we study the confining theories and
their responses to the external electric field. 
{The confining theory is formulated by the bulk solution which is obtained
by retaining the dilaton and the axion. In the framework of this model, it
is possible to introduce the temperature by adopting the AdS-Schwartzschild
like solution \cite{GSUY}. However, the transition point from the deconfinement
to confinement is at the zero temperature in this theory. So, in the 
confining phase, we can not give a model with horizon corresponding to the
low temperature where the confinement is retained. It is however possible
to consider a finite temperature model without the horizon when we 
artificially put an upper bound for the time and impose a periodic condition
in the Euclidean metric. In this case, we could find the confinement-
deconfinement transition temperature as the Hawking-Page transition \cite{ET}.  
The same kind of phase-transition can be seen also in the type IIA models as 
\cite{KMMW2,HT}.

Here we, however, consider the zero-temperature confining theories.}
In this case, we can not introduce a finite $n_b$
since there is no horizon as in the high temperature phase.
As a result of this fact, the conductor phase is not found for small electric field.
However, we could find the insulator-conductor transition 
when the electric field goes over
a critical value ($E_{cr}=$ the tension of the linear potential between the quark
and the anti-quark), 
where the repulsion due to the electric field exceeds
the attractive confining force between the quark and the anti-quark.
This is consistent with the result $E_{cr}=0$ in the deconfinement theory since
there is no (long range) confining force competing with the repulsion due to
the finite electric field. 

The existence of such conductor phase in the confining phase
implies that there must be the carrier of this current.
In the confining theory, it may be possible to
create pairs of baryon and anti-baryon as baryon number carriers \cite{BLL}.
On the other hand, the pair creation of quark and anti-quark seems to be impossible 
since they should be bounded to mesons. 


However, we could consider here that the carriers of this current are the quark 
and the anti-quark created by 
the meson melting. This is assured by the existence of
the quasi-normal modes of mesons, which
should decay with a definite life-time.
In the conductor phase, we actually find the quasi-normal modes of mesons
in terms of the embedded D7 brane as in the high temperature phase
\cite{Starinets,Hoyos,Kovtun,Shock,evans}. 
In this case, instead of the black-hole horizon, the incoming
wave of mesons are absorbed into the "locus vanishing" point ($r^*$), and the mesons
are broken into quark, anti-quark and partons with radiations which are emitted
from the accelerated charged particles due to the strong external electric field
($E>E_{cr}$). 
However the whole configuration of a quark and an anti-quark
,in this case, would be a string connecting the probe D-brane(s).
This configuration would be changed to the strings
connecting the Rindler horizon and probe D-brane after an appropriate
coordinate transformation \cite{Xiao}. 
More on this point, we discuss in the section 4.
\begin{figure}[htbp]
\vspace{.3cm}
\begin{center}
\includegraphics[width=6cm]{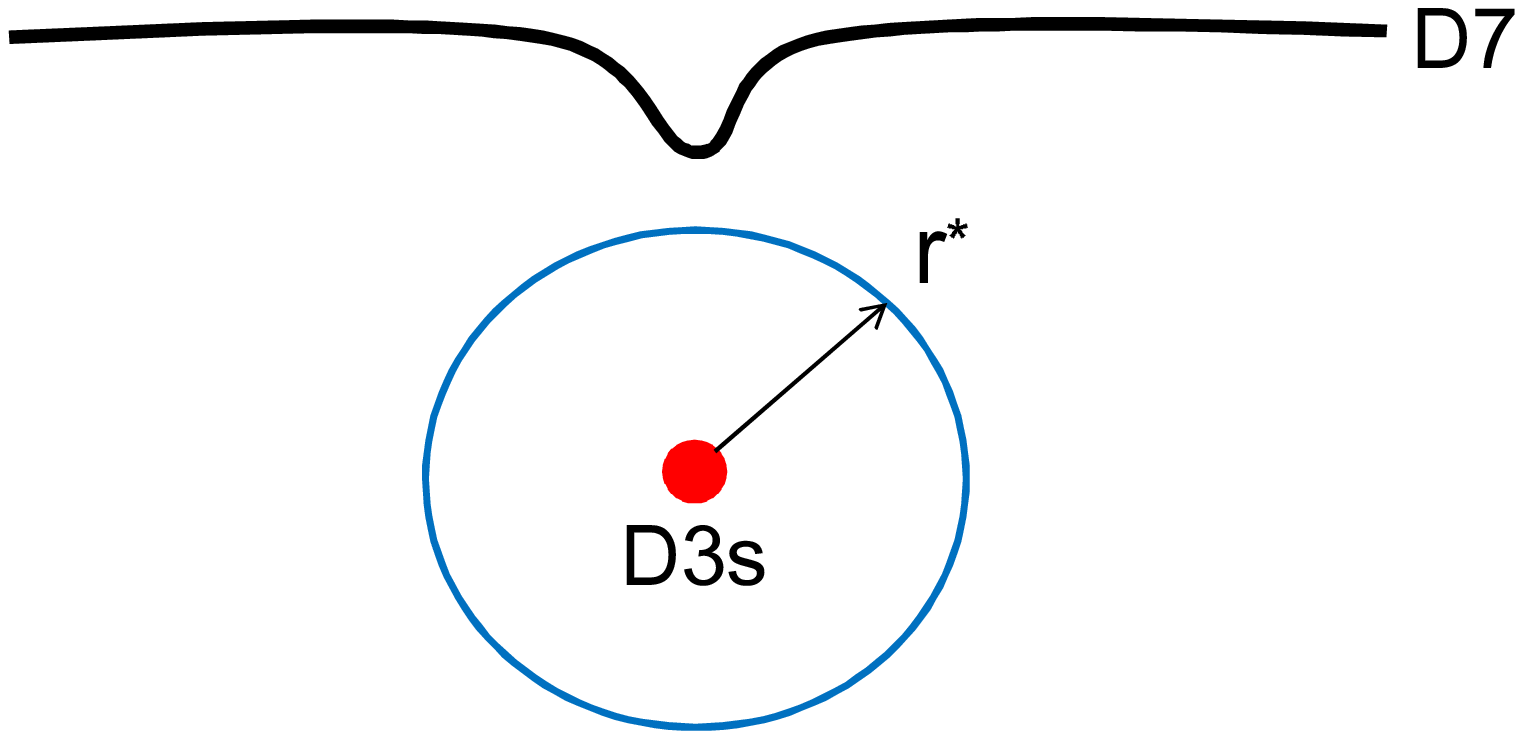}
\includegraphics[width=6cm]{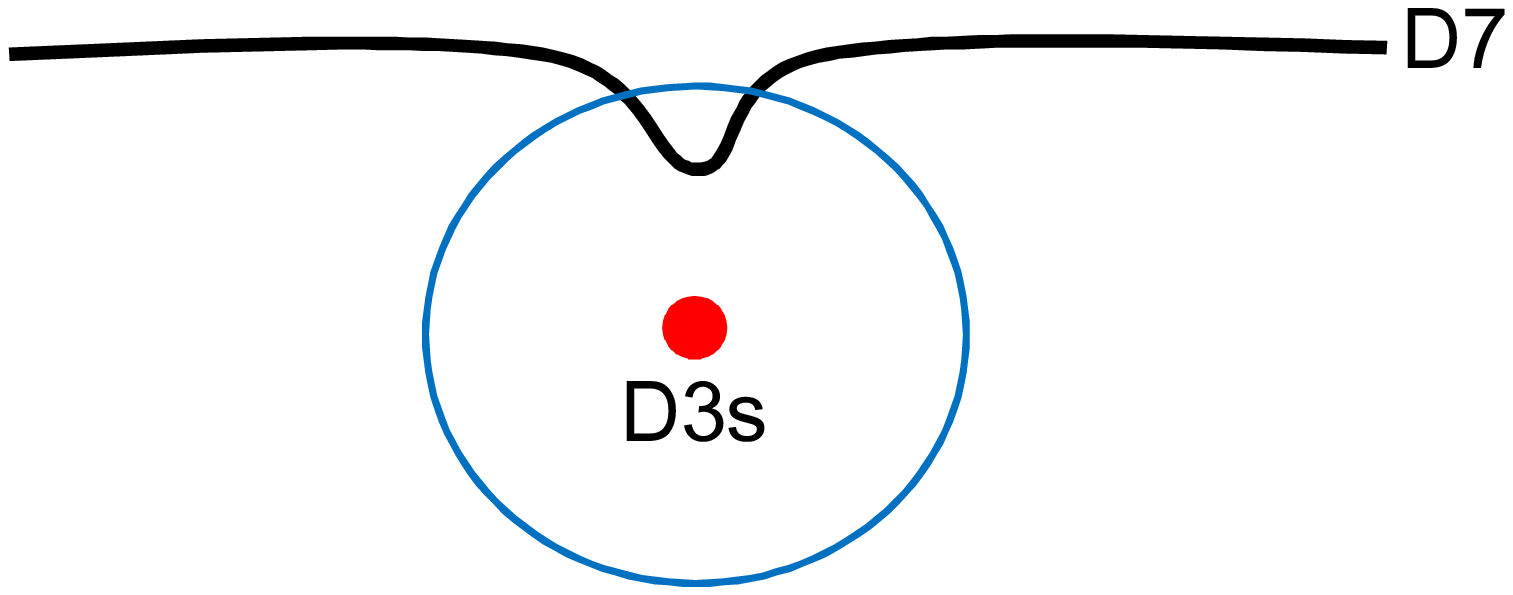}
\caption{\small Two typical D7 brane embeddings in the bulk space-time 
corresponding to a supersymmetric confining 
theory. The left and right hand figures show the insulator and the conductor
phase respectively. The radius $r^*$ is explained in the text.}
\label{sch1}
\end{center}
\end{figure}

\begin{figure}[htbp]
\vspace{.3cm}
\begin{center}
\includegraphics[width=6cm]{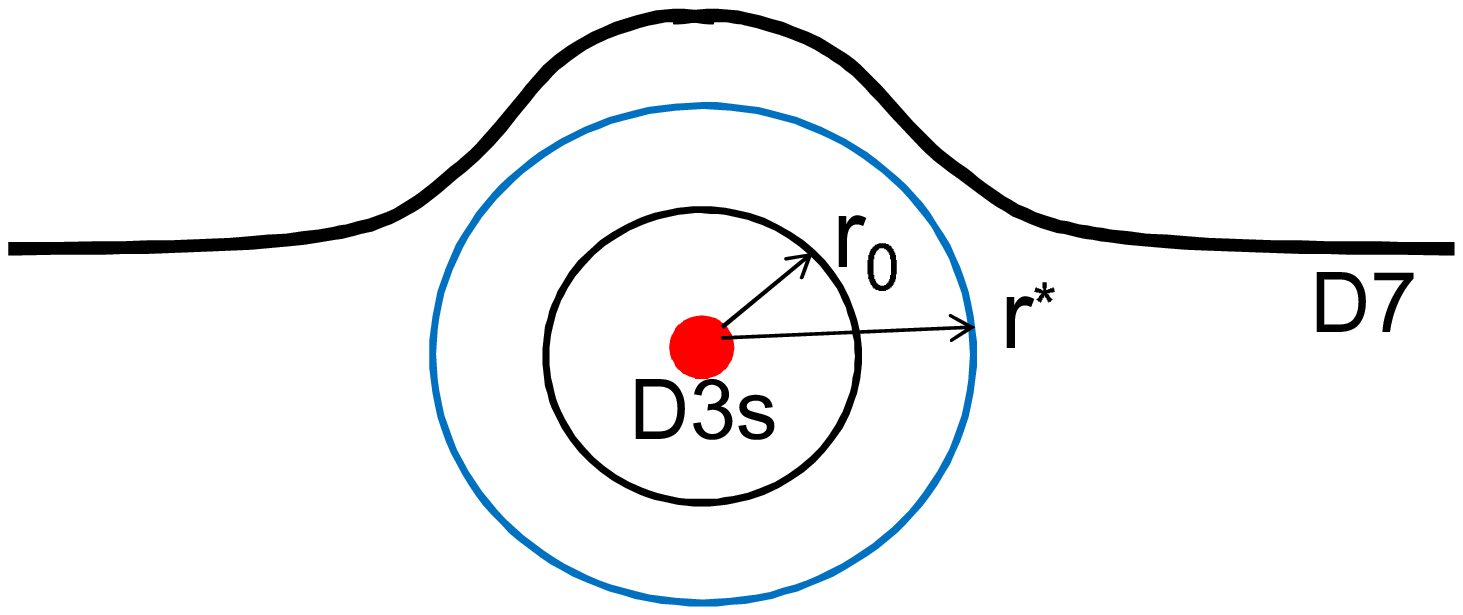}
\includegraphics[width=6cm]{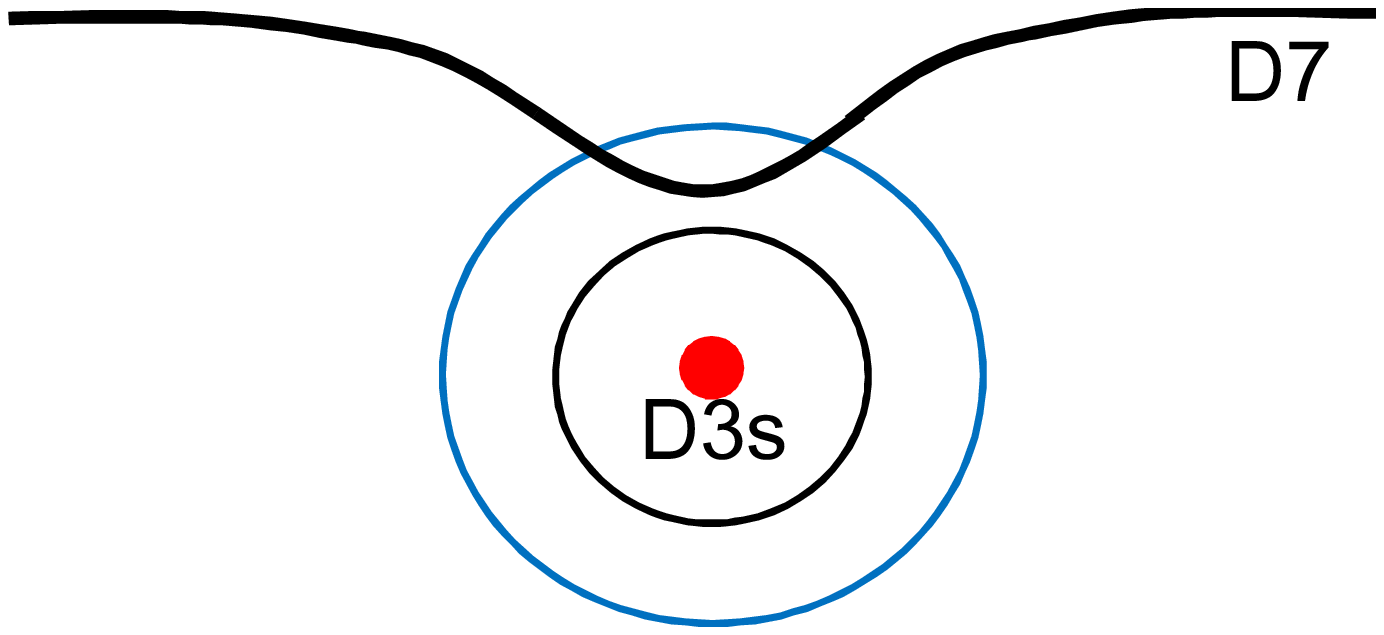}
\caption{\small The D7 embeddings for non-supersymmetric confining theory are shown.
The left (right) shows insulator (conductor) phase. In this case, singularity 
exists and its position is shown
by the radius $r_0$.}
\label{sch2}
\end{center}
\end{figure}

Other than the insulator-conductor transition, we find two new phase 
transitions in the confining theories. One is found in the insulator phase
of the confining theory with chiral and super symmetries. For a fixed $E(<E_{\rm cr})$,
we find a jump of the chiral condensate and also of the meson mass for light quarks
$m_q(<m_q^{\rm cr})$, where $m_q^{\rm cr}$ varies with $E$ and we show a phase
diagram in $m_q$-$E$ plane. {We will see that the
phase is due to the confining configuration.}
Another is the chiral phase transition in the confining theory with
broken chiral and super symmetries. We find the vanishing of the chiral
condensate for the massless quark at the insulator-conductor transition
point. This implies the restoration of the chiral symmetry for
$E>E_{\rm cr}$,

The pictures of D7 brane embeddings are shown for the two confining theories in the 
Figs. \ref{sch1} and \ref{sch2}. In the Fig. \ref{sch1}, supersymmetric case is shown,
and the non-supersymmetric one is given in the Fig. \ref{sch2}. The left hand figures
of them show the insulator phase, and the chiral symmetry is preserved (broken) in 
the case of Fig.\ref{sch1} (Fig. \ref{sch2}). In the conductor phase, shown by the
right hand figures, the chiral symmetry is preserved in both cases.


\vspace{.2cm}
In section 2, we give the setting of our model for the supersymmetric 
and non-supersymmetric version
of confining Yang-Mills theory. And the response to the electric field is 
studied in the following sections 3 and 4 for the supersymmetric case. 
For non-supersymmetric case, the analysis including the chiral transition
are given in the section 5. and
the summary is given in the final section.

\section{D3/D7 model for confining YM theory}

We start from
10d IIB model retaining the dilaton
$\Phi$, axion $\chi$ and self-dual five form field strength $F_{(5)}$.
Under the Freund-Rubin
ansatz for $F_{(5)}$, 
$F_{\mu_1\cdots\mu_5}=-\sqrt{\Lambda}/2~\epsilon_{\mu_1\cdots\mu_5}$ 
\cite{KS2,LT}, and for the 10d metric as $M_5\times S^5$ or
$ds^2=g_{MN}dx^Mdx^N+g_{ij}dx^idx^j$, we find the solution.
The five dimensional $M_5$ part of the
solution is obtained by solving the following reduced 5d action,
\beq\label{10d-action}
 S={1\over 2\kappa^2}\int d^5x\sqrt{-g}\left(R+3\Lambda-
{1\over 2}(\partial \Phi)^2+{1\over 2}e^{2\Phi}(\partial \chi)^2
\right), \label{5d-action}
\eeq
which is written 
in the string frame and taking $\alpha'=g_s=1$. 

\vspace{.3cm}
The solution is obtained under the ansatz,
\beq
\chi=-e^{-\Phi}+\chi_0 \ ,
\label{super}
\eeq
which is necessary to obtain supersymmetric solutions. 
And the solution is expressed as
$$ 
ds^2_{10}=G_{MN}dX^{M}dX^{N} ~~~~~~~~~~~~~~\qquad
$$ 
\beq\label{background}
=e^{\Phi/2}
\left\{
{r^2 \over R^2}A^2(r)\left(-dt^2+(dx^i)^2\right)+
\frac{R^2}{r^2} dr^2+R^2 d\Omega_5^2 \right\} \ . 
\label{finite-c-sol}
\eeq 

Then, the supersymmetric solution is obtained as
\beq
e^\Phi= 1+\frac{q}{r^4}\ , \quad A=1\, ,
\label{dilaton}
\eeq
where $M,~N=0\sim 9$ and
$R=\sqrt{\Lambda}/2=(4 \pi N)^{1/4}$. 
And $q$ represents the vacuum expectation value (VEV) 
of gauge fields condensate~\cite{GY}. 
In this configuration, the four dimensional boundary represents the 
$\cal{N}$=2 SYM theory. In this model, we find quark confinement in the
sense that we find a linear rising potential between quark and anti-quark
with the tension $\sqrt{q}/R^2$ \cite{KS2,GY}.

\vspace{.3cm}
As for the non-supersymmetric case, the solution is given by (\ref{background})
and
\begin{equation}\label{non-susy-sol}
A(r)=\left((1-(\frac{r_0}{r})^8)\right)^{1/4},\qquad
e^{\Phi}=\left(\frac{(r/r_0)^4+1}{(r/r_0)^4-1}\right)^{\sqrt{3/2}},\qquad
\chi=0\, .
\end{equation}
This configuration has a singularity at the horizon $r=r_0$. 
So we can not extend our analysis to near this horizon where higher curvature
contributions are important.
This theory provides confinement and chiral symmetry breaking. The latter
means that we find non-zero chiral condensate for the massless quark. In 
other
words, a dynamical quark mass would be generated for a massless quark in 
this
theory. This point is different from the above supersymmetric background 
solution.
The confinement is sustained by the gauge condensate, which is proportional
to $r_0^4$ in the present case \footnote{This point is easily assured by expanding
$e^{\Phi}$ in (\ref{non-susy-sol}) by the powers of $r_0/r$.
}, as in the supersymmetric case.

\vspace{.3cm}
\section
{\bf D7 brane embedding and phase transitions}~
The D7 brane is embedded in the above background (\ref{background}) as follows. 
First,
the extra six dimensional part of the above metric (\ref{finite-c-sol})
is rewritten as,
\beq
 \frac{R^2}{r^2} dr^2+R^2 d\Omega_5^2
 =\frac{R^2}{r^2}\left(d\rho^2+\rho^2d\Omega_3^2+(dX^8)^2+(dX^9)^2
\right)\ ,
\eeq
where $r^2=\rho^2+(X^8)^2+(X^9)^2$.
And we obtain the induced metric for D7 brane,
$$ 
ds^2_8=e^{\Phi/2}
\left\{
{r^2 \over R^2}A^2\left(-dt^2+(dx^i)^2\right)+\right.
\qquad
$$
\beq
\left.\frac{R^2}{r^2}\left((1+(\partial_{\rho}w)^2)d\rho^2+\rho^2d\Omega_3^2\right)
 \right\} \ , 
\label{D7-metric}
\eeq
where we set as $X^8=w(\rho)$ and $X^9=0$
without loss of generality due to the rotational invariance in
$X^8$-$X^9$ plane. The embedded configuration is obtained as the solution
for the profile function $w(\rho)$, and it is performed with non-trivial
gauge field 
\beq\label{electric-f}
A_x(\rho,t)=-Et+h(\rho)
\eeq
in the D7 brane action. Here the chemical potential and the charge density are not
introduced since we are considering in confinement phase where no
free quark is allowed.

\vspace{.3cm}
The brane action for the D7-probe is given as
$$ 
S_{\rm D7}= -\tau_7 \int d^8\xi \left(e^{-\Phi}
    \sqrt{-\det\left({\cal G}_{ab}+2\pi\alpha' F_{ab}\right)}
      -{1\over 8!}\epsilon^{i_1\cdots i_8}A_{i_1\cdots i_8}\right)
$$
\beq
   +\frac{(2\pi\alpha')^2}{2} \tau_7\int P[C^{(4)}] \wedge F \wedge F\ ,
\label{D7-action}
\eeq
where $F_{ab}=\partial_aA_b-\partial_bA_a$.
${\cal G}_{ab}= \partial_{\xi^a} X^M\partial_{\xi^b} X^N G_{MN}~(a,~b=0\sim 7)$
and $\tau_7=[(2\pi)^7g_s~\alpha'~^4]^{-1}$ represent the induced metric and
the tension of D7 brane respectively.
And $P[C^{(4)}]$ denotes the pullback of a bulk four form potential,
\beq
C^{(4)} = 
\left(\frac{r^4}{R^4} d x^0\wedge d x^1\wedge
d x^2 \wedge d x^3 \right)\ .
\label{c4}
\eeq
The eight form potential $A_{i_1\cdots i_8}$,
which is the Hodge dual to the axion, couples to the
D7 brane minimally. In terms of the Hodge dual field strength,
$F_{(9)}=dA_{(8)}$ \cite{GGP}, the potential $A_{(8)}$ is obtained. 

\vspace{.3cm}
Then, by taking the canonical gauge, we arrive at the following D7 brane
action,
\beq
S_{\rm D7} =-2\pi^2\tau_7~\int d^4x d{\rho}\rho^3 \left({R\over r}
   e^{\Phi/2}\sqrt{P e^{\Phi}-Q}-{q\over r^4} \right)
\ ,
\label{D7-action-2}
\eeq
\beq
 P=|\tilde{G}_{00}|\tilde{G}_{xx}\tilde{G}_{\rho\rho}\, ,
 \quad Q=\tilde{G}_{\rho\rho}{\dot{\tilde{A}}_x}^2
-|\tilde{G}_{00}|{{\tilde{A'}}_x}^2
\eeq
where $\tilde{G}_{MN}=e^{-\Phi/2}{G}_{MN}$, 
$\tilde{A}_{x}=2\pi\alpha'{A}_{x}$ and the eight form part
is given as $C_8(r)={q/r^4}$ \cite{GIN}. The explicit forms of $\tilde{G}_{MN}$
are given as follows,
\beq
 |\tilde{G}_{00}|=\tilde{G}_{xx}=\left({r\over R}\right)^2\, , \quad
 \tilde{G}_{\rho\rho}=\left({R\over r}\right)^2(1+w'(\rho)^2).
\eeq
At first, we solve the equation of motion of ${\tilde{A'}}_x=\tilde{h'}$ as
\beq\label{electric}
  e^{\Phi/2}{\rho^3\over r}{|\tilde{G}_{00}|\tilde{A'}_x
\over \sqrt{P e^{\Phi}-Q}}=B
\eeq
where $B$ denotes a constant and it corresponds to the electric current,
\beq
  \langle J_x\rangle=B\, .
\eeq

Then we rewrite the D7 action by eliminating $\tilde{A'}_x$ through
the Legendre transformation,
\beq
 U\equiv S-\int d^4x d\rho \tilde{A'}_x {\delta S\over \delta \tilde{A'}_x}
\eeq
and we obtain
\beq
  U=-2\pi^2\tau_7~\int d^4x d{\rho} L
\eeq
\beq
  L=\sqrt{{\tilde{G}_{\rho\rho}\over |\tilde{G}_{00}|}}
  \left\{\left(e^{\Phi} |\tilde{G}_{00}|\tilde{G}_{xx} -\tilde{E}^2\right)
  \left({\rho^6\over r^2} e^{\Phi}|\tilde{G}_{00}|
     - B^2\right)\right\}^{1/2}-{q\over r^4}{\rho^3\over R}
\eeq
Explicit expression of $L$ in the present case is given as follows,
\beq
  L=\left({R\over r}\right)^2\sqrt{1+{w'}^2}\sqrt{\left({q+r^4\over R^4}
-\tilde{E}^2\right)\left({\rho^6 \over R^2}e^{\Phi}-B^2\right)}
-{q\over r^4}{\rho^3\over R}
\eeq
Then in our case, the locus vanishing
point, where the insulator and conducting phase
is separated, is given as
\beq
  {r^*}^2={\rho^*}^2+w^2(\rho^*)=\sqrt{\tilde{E}^2R^4-q}
\eeq
which is shifted by $q$. In other words, there is a threshold of $\tilde{E}$
at $q^{1/2}/R^2$ to generate a locus or the conducting area. Namely, for 
$\tilde{E}<q^{1/2}/R^2$, the electric current $B$ should be zero.

The equation of motion of $w(\rho)$ is obtained from the above $L$ as follows,
\beq\label{profile-eq}
 \partial_{\rho}\left({w'\over \sqrt{1+{w'}^2}}\sqrt{F(r)}\right)
 -{w\over r}\left({F'(r)\over 2\sqrt{F(r)}} \sqrt{1+{w'}^2}
 +{4q\rho^3\over r^5}\right)=0\, ,
\eeq
\beq
 F(r)={\rho^6}e^{2\Phi}\left(1-{R^4\tilde{E}^2\over r^4e^{\Phi}}\right)
\left(1-{R^2B^2\over \rho^6e^{\Phi}}\right)\, ,
\eeq
where $F'(r)=\partial_rF(r)$.

\vspace{.3cm}
\subsection{Insulator Phase}

\begin{figure}[htbp]
\vspace{.3cm}
\begin{center}
 \includegraphics[height=6cm,width=6.5cm]{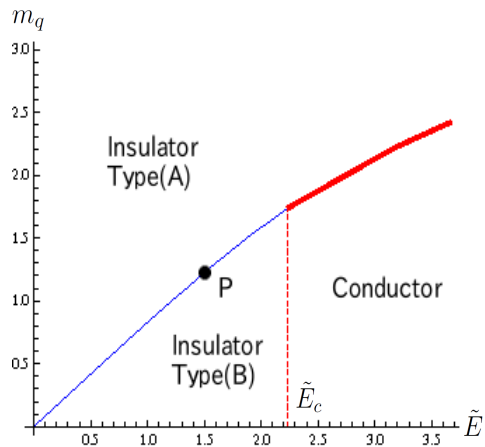}
\caption{{\small The phase diagram in the $m_q$-$\tilde{E}$ space at $q=5$ and $R=1$.
The thick solid (red) curve and the dashed (red) vertical line ($\tilde{E_c}=\sqrt{q}/R^2$)
represent critical points
 between the insulator phase and the conductor phase. The thin solid
 (blue) curve represents the critical points between type (A) and type (B)
phases in the same insulator phase. The point P corresponds to a
 transition point, and the jump
of $c$ on this point is shown in the Fig. \ref{ww}.
 \label{phase-diagram}}}
\end{center}
\end{figure}

Here we concentrate on the insulator phase, where the electric field is 
restricted as
\beq\label{force-1}
  {q\over R^4}\geq\tilde{E}^2\, .
\eeq
In this case, we find $\left({q+r^4\over R^4}-\tilde{E}^2\right)>0$ for any $r$, then
we could set $B=0$ or $\langle J_x\rangle=0$
in order to keep $F(r)$ to be positive. Then in this phase,
the electric current is zero and we can call this case
as insulator (see Fig. \ref{phase-diagram}).
The inequality (\ref{force-1}) implies that the repulsive force between the quark and anti-quark
due to the electric field is smaller than the attractive color force to confine
them. For $q\neq 0$, we find the tension between the quark and anti-quark is equal to $q^{1/2}/R^2$ \cite {GY}.
Thus, we can understand the above statement. 

\vspace{.3cm}
Under this setting, we solve the equation (\ref{profile-eq})
and we could find two phases, type (A) and (B), which are shown in the 
phase diagram of the Fig. \ref{phase-diagram}.
The asymptotic form of the solution 
at $\rho\to \infty$ is obtained as
\beq
w(\rho)=m_q+\frac{c}{\rho^2}+\cdots\, ,
\eeq
where $m_q$ and $c=\langle \bar{\psi}\psi\rangle$ corresponds to
the quark mass and the chiral condensate respectively.
This correspondence is well known from the AdS/CFT dictionary. These quantities
are fixed after solving (\ref{profile-eq}) in all region of $\rho$ including
the infrared region of $\rho=0$. The value of $c$ for any solution depends on
$m_q$, $q$ and also $E$.

\vspace{.3cm}
For $q=0$, the background is reduced to the supersymmetric AdS$_5\times S^5$ and 
we find $c=0$ when $E=0$. 
When $E$ is turned on,
we find the solution of negative $c$, which depends on the quark mass $m_q$
and decreases with decreasing $m_q$ monotonically. 
This case has been studied in \cite{AFJK2}.

\begin{figure}[htbp]
\vspace{.3cm}
\begin{center}
 \includegraphics[width=7cm]{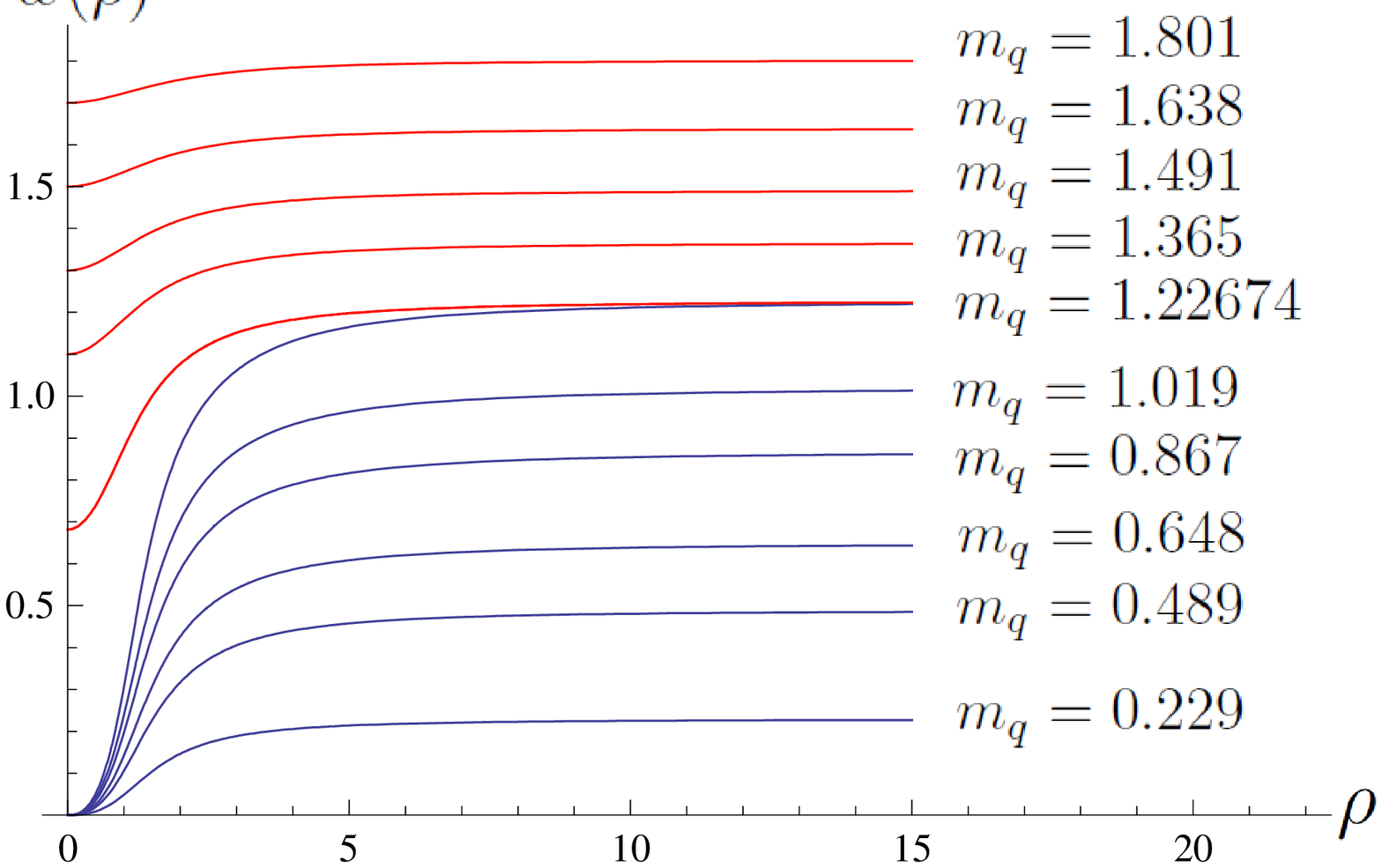}
 \includegraphics[width=7cm]{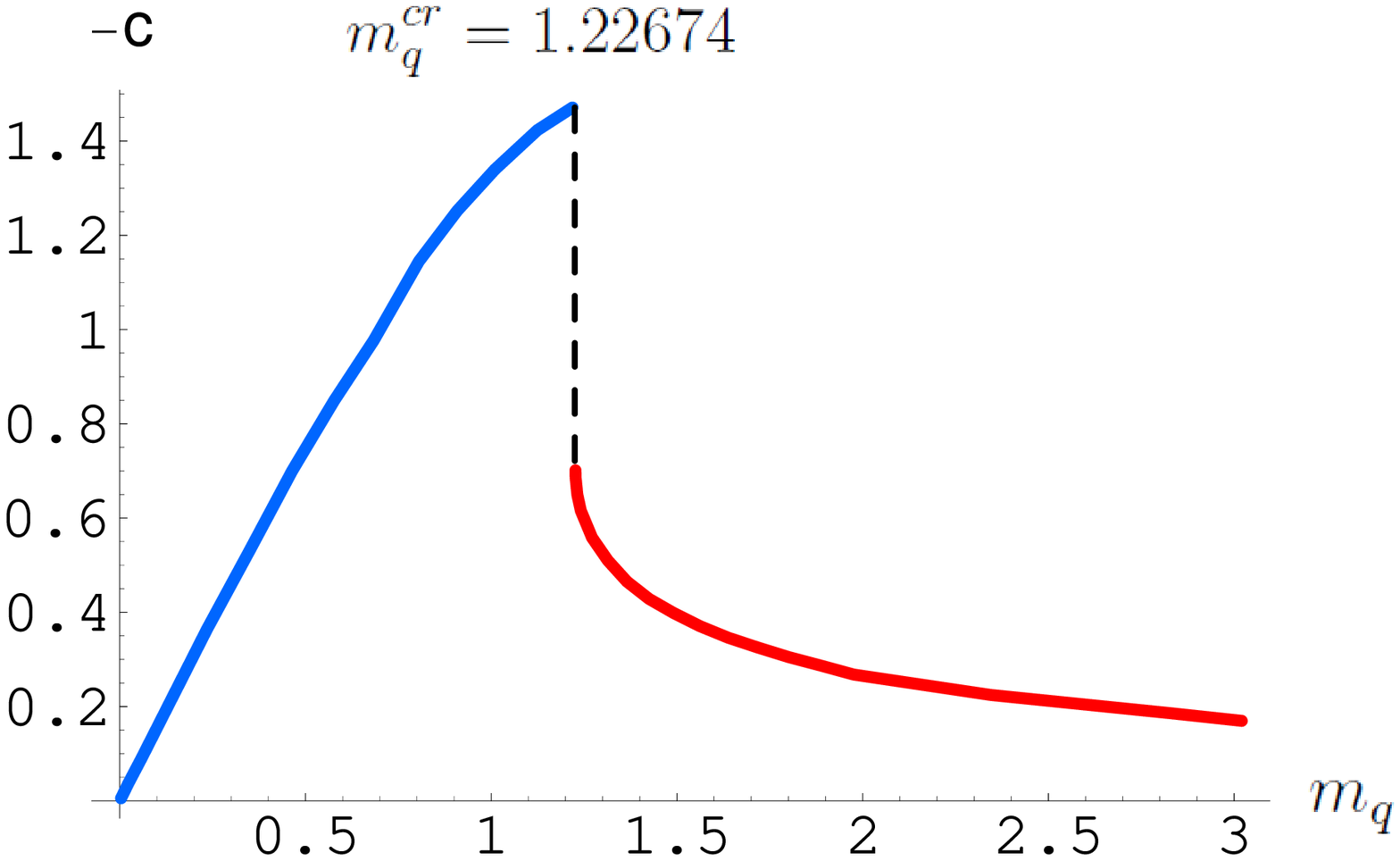}
\caption{{\small Solutions $w(\rho)$ and $c$ for the various $m_q$ at $\tilde{E}=1.5$, $q=5$ and $R=1$. 
The left figure shows that solutions become almost zero 
at the $\rho=0$ for $m_q<m_q^{cr}(=1.22674)$, 
on the other hand, solutions are always finite for $m_q>m_q^{cr}$. 
The right figure shows that the chiral condensates $c$ jump at
 $m_q^{cr}$. }}
\label{ww}
\end{center}
\end{figure}

Here we concentrate on the case of $q>0$, and the $m_q$ dependence of $c$ 
has been studied. For the case of $\tilde{E}=1.5$, $q=5$ and $R=1$,
the D7 embedded solutions and the value of $c$ for various $m_q$ are shown
in the Fig. \ref{ww} as a typical example. This analysis has been performed
across the point $P$ in the Fig. \ref{phase-diagram}.
From the Fig. \ref{ww},
a phase transition is observed through a jump of $c$ at $m_q=m_q^{\rm cr}\simeq 1.23$.
This transition point
depends on other parameters of the theory, especially sensitive to $E$.
In the Fig \ref{phase-diagram}, the line of this phase transition is
shown in the $m_q-E$ plane. Each phase is assigned as (A) and (B) in the Fig. \ref{phase-diagram}.

\vspace{.3cm}
The jump of $c$ is explicitly reflected to the jump of the infrared end point
of the solution, $w(0)$, which changes from a finite value to the very small
value (almost zero) for the small  
quark mass, namely
in the region $0\leq m_q\leq m_q^{\rm cr}$. 
On the other hand,
it is known that the meson mass is proportional to $w(0)$ in the supersymmetric
case of $E=0$, and we could see that
this relation is retained even if $E>0$. Namely, after this transition
for small $m_q$, the meson mass of $m_q\leq m_q^{\rm cr}$ also jumps from a finite value to 
almost zero.

\vspace{.3cm}
Finally we give a comment about the similarity of the role of electric field $E$
to the temperature.
The dynamical situation observed here is very similar to the case of the high
temperature phase. 
The temperature gives thermal screening to the system of confining quarks to suppress
the color force. The electric field $E$, on the other hand, 
pulls the quark and anti-quark to the opposite direction
to compete the attractive confining force between the quark and anti-quark. 
In both cases, the confining force is effectively suppressed.
In this sense,
the role of $E$ is similar the temperature. As a result of this effect,
$E$ pulls down the end point of D7 toward $r=0$, the horizon of the background
metric. Our model gives quark confinement for $E=0$ 
due to the parameter $q$ \cite{GY}, and the confining force is still stronger
than the repulsion of $E$ in the phase (B). However we could find
the conducting phase for $\tilde{E}\geq \sqrt{q}/R^2$ where the confinement force
is defeated by the repulsion due to $E$ as shown below.


\subsection{Meson spectrum }

Here we consider meson spectrum in the insulator phase
to understand well the phase transition observed in the previous section.
The meson masses are given by solving the equations of motion of fluctuations
of the fields on the D7 brane. The simplest and non-trivial one are the fluctuations
of the gauge fields which are perpendicular to $(t,x,\rho)$ directions. They 
do not mix with the other fluctuations and are
denoted by $A_{\bot}$. 
We impose the following ansatz for $A_{\bot}$,
\beq
\delta A_{2,3}=\delta A_{\bot}= e^{-i(\omega t -k_x x)}A_{\bot}(\rho)\, ,    
\eeq
where we consider the $y, z$ components of $\delta A_{\bot}=\delta A_{y,z}$ for 
the simplicity.
Then, the action is expanded up to the quadratic part about $\epsilon$,
\bea \label{meson}
S_{D7}
&=&\tau_7\int d^8\xi L_0\left(1+\tilde{L}^{(2)}_{A_{\bot}}+\cdots\right)\, , \nonumber \\
     L^{(2)}_{A_{\bot}}&{\equiv}& L_0\tilde{L}^{(2)}_{A_{\bot}}=
{1\over 2}L_0\tilde{G}^{ab}\tilde{G}^{\bot\bot}
     \partial_a\delta A_{\bot}\partial_b\delta A_{\bot}\, , 
\eea
where
\bea\label{inverse-metric}
\tilde{G}^{\rho\rho}&=&\left({r^4\over R^4}-\tilde{E}^2e^{-\Phi}\right)
/G_{(3)}\, , \\
\tilde{G}^{\rho t}&=&-{{B}\tilde{E}R^2\over \rho^3 re^{\Phi}\sqrt{G_{(3)}}}\, , \\
\tilde{G}^{tt}&=&-\left(1+{w'}^2+{B}^2
  {G_{(3)}R^4\over \rho^6 r^2e^{\Phi}}\right)/G_{(3)}\, , \\
 \tilde{G}^{\bot\bot} &=&\left({r\over R}\right)^2\, , \\
G_{(3)}&=&{r^2\rho^6\over R^4}(1+w'(\rho)^2){1-\frac{R^4\tilde{E}^2}{r^4e^{\Phi}} 
\over {\rho^6/ R^2}-{{B}^2 e^{-\Phi}} }\, ,
\eea

\beq
  L_0={\rho^6\over R}\sqrt{(1+w'(\rho)^2)}\sqrt{\left(1-\frac{R^4\tilde{E}^2}{e^{\Phi}r^4}\right)/ \left( {\rho^6\over R^2}-\frac{B^2}{e^{\Phi}}\right)}\, .
\eeq
In the case of the insulator phase, we set as $B=0$, then we obtain
\beq
L^{(2)}_{A_{\bot}}= e^{-2i(\omega t-k_x x)}\rho^3~\frac{-\frac{R^4}{r^4}(1+w'(\rho)^2)M^2A_{\bot}^2(\rho)+\left(1-\frac{R^4\tilde{E}^2}{e^{\Phi}r^4}\right) {A'}_{\bot}^2(\rho)}{\sqrt{(1+w'(\rho)^2)\left(1-\frac{R^4\tilde{E}^2}{e^{\Phi}r^4}\right)}}\, ,
\eeq 
where prime denotes the derivative with respect to $\rho$ and
$M^2=-k^2=-\omega^2+k_x^2$, which defines the mass of the mesons.
Then the equation of motion of $A_{\bot}(\rho)$ is obtained as follows, 
\begin{eqnarray}\label{meson-mass}
A''_{\bot}(\rho)+\left({3\over \rho}+{H'\over H}\right)A'_{\bot}(\rho)
  +\frac{R^4M^2}{r^4H^2(\rho)}A_{\bot}(\rho)=0 ,\\
H(\rho)=\sqrt{\frac{1-\frac{R^4\tilde{E}^2}{e^{\Phi}r^4}}{1+w'(\rho)^2}} \ .
\end{eqnarray}

\begin{figure}[htbp]
\vspace{.3cm}
\begin{center}
\includegraphics[height=5.5cm,width=7cm]{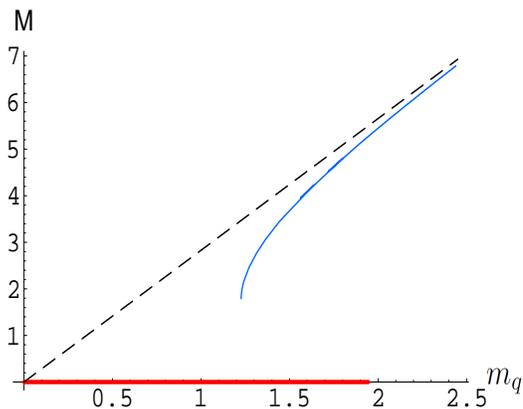}
\caption{{\small The lightest meson mass for $\tilde{E}=1.5$, $q=5$ and $R=1$. 
The (blue) solid curve shows the meson mass of Minkowski embedding solutions. 
The (red) horizontal line shows (almost) massless mesons of the solution of embedding at the origin at the end point.
The dashed line represents the lightest meson for $\tilde{E}=0$ given by (\ref{meson-mass-2}) since $m_q=w(0)$. 
For $m_q<m_{cr}=1.22674$, all meson masses are almost zero.}}
\label{ms}
\end{center}
\end{figure}

Here we notice the following two points. 
Firstly, the mass eigenfunction should
be normalizable with respect to $\rho$ integration in (\ref{meson}), then 
we impose the following boundary condition,
\beq
  \int^{\infty} d\rho L^{(2)}_{A_{\bot}}<\infty\, .
\eeq
Then we find the following condition for the asymptotic behavior of $A_{\bot}(\rho)$
at large $\rho$,
\beq
A_{\bot}(\rho)|_{\rho\to\infty}\to O(\rho^{-\beta}) \to 0\, ,
\quad \beta>1\, .
\eeq
This condition leads to discrete meson mass spectrum. 

\vspace{.3cm}
Secondly, consider
the infrared limit ($\rho=0$) for the equation (\ref{meson-mass}). Assuming the 
leading term of $A_{\bot}$ in this region as
\beq
 A_{\bot}(\rho)|_{\rho\to 0}\simeq \rho^{\alpha}\, .
\eeq
Then, the asymptotic form of the equation is considered by separating
to the two cases. 

(i) For $w(0)\neq 0 (>0)$, we find from (\ref{meson-mass}) the following form,
\beq\label{finite-w}
  \alpha(\alpha+2)\rho^{\alpha-2}+\frac{R^4M^2}{w(0)^4H^2(0)}\rho^{\alpha}
       +\cdots=0
\eeq
where $\cdots$ denotes the terms of higher order of $\rho$. 

(ii) For $w(0)=0$, we have
\beq\label{zero-w}
  \frac{R^4M^2}{H^2(0)}\rho^{\alpha-4}+\alpha(\alpha+2)\rho^{\alpha-2}
       +\cdots=0\, .
\eeq

\vspace{.3cm}
For the first case, (i) $w(0)\neq 0$, we find $\alpha=0$ or $-2$ from (\ref{finite-w}).
And we find finally $\alpha=0$ from the normalizability of $A_{\bot}(\rho)$.
Then, finite mass eigenvalues of $M$ are found by these boundary conditions in this case.

As for the 
second case of $w(0)=0$, we must take $M=0$ and $\alpha=0$. This implies that
all the
meson mass of the bound state of quark and anti-quark with $m_q\leq m_q^{\rm cr}$
should be zero.

This is consistent with the meson spectra obtained for $\tilde{E}=0$ \cite{KMMW},
\beq\label{meson-mass-2}
M=2\sqrt{n(n+1)}\frac{w(0)}{R^4}\, , \quad n=1,2,\cdots.    
\eeq 
And also at finite $E$, we find approximately this formula for the meson spectrum.
The numerical results are shown in the Fig. \ref{ms} for $\tilde{E}=1.5$ and $q=5$. This
result supports our statement given above.

As for the exact value of $w(0)$ in this case, it is difficult to say that
it is zero or finite but small
since there is no analytical support of this statement. So, we can say only
that the meson mass is very small in the phase (B).

\section{Insulator-Conductor transition}
Here we consider the large $E$ which satisfies the following inequality, 
\beq\label{force2}
\tilde{E}^2>{q\over R^4}\, .
\eeq 
In this case, there is a position 
of $r$ called as a locus vanishing point defined as
\beq
  r^*=\left(R^4\tilde{E}^2-q\right)^{1/4}\, ,
\eeq
where the action vanishes. Then
we need non-zero $B$ in order to preserve the reality of the action
even in the region of $r<r^*$ \cite{Karch:2007pd}. 
This guarantees the existence of
the solution $w(\rho)$ up to the region, $w^2(\rho)<{r^*}^2-\rho^2$, since
the equation of motion of $w$ must be also real.
Then, for these solutions, $w$, the electric current $B$ must be accompanied
in spite of the fact that the theory
is in the quark confinement phase, where we can not introduce charge density
$n_b$ as mentioned above.

\vspace{.3cm}
\noindent {\bf Carrier of the current $B$:}

Up to now, this phenomenon has been considered
in the non-confining high temperature phase, 
and the carrier of this current can be identified with the 
quark and the anti-quark which are pair created  
by the strong electric filed \cite{Karch:2007pd}. 
And, the configuration of
the quark or the anti-quark is given by the string
connecting the probe brane and the event horizon, which exists in the 
holographic high temperature model. 

In our confining theory, however such a quark string configuration 
can not be considered
since the event horizon does not exist.
One may therefore wonder what is the carrier
of this baryon number current, $B$, in the present confining case. 

One possibility might be the pair created
baryons and anti-baryons which are allowed in the confinement phase
{as considered in the type IIA model \cite{BLL}. Another possibility is to
consider the creation of mesons which have opposite charge. But the latter case
would be meaningful only for some flavor non-singlet current as discussed in \cite{Erdmenger:2007bn}. In the present case, however,
we restrict our attention to the $U(1)_B$ current,
so we don't consider this case. 

The first possibility
does not contradict with the present model.
But, in our model, the current
appears when the electric force ($\tilde{E}$)
for a unit quark number exceeds the confinement
attractive force, which is given by the tension,
$q^{1/2}/R^2$, of linear rising potential between the quark and
the anti-quark to bind them \cite{GY}. 
Therefore, we are led to the idea that the 
main part of the current should be reduced
to the pair created quark and anti-quark, which can not be bound to mesons since
the attractive force is suppressed by the strong electric repulsion. 

Meanwhile, we must remind that the theory considered here describes the confining
phase. Then, it seems to be controversial to suppose the quark and the anti-quark 
as the current carrier. In spite of this fact,}
as explained below, it would be natural to consider such that the carriers of the 
current $B$ would be the quarks and the anti-quarks. One reason is that
the mesons would be melt down into quarks and anti-quarks for the solutions ($w$)
with finite $B$ as shown in the next subsection
for the case of quarks with small-mass compared to the given $\tilde{E}$.

However we should notice that the melted mesons may not be changed to the 
string state of quarks as seen
in the high temperature deconfinement phase. 
The reason is that the dual gravitational geometry 
of the confining gauge theories given here has no event horizon, 
which could be the end point of the free quark-string configuration.
{As the resolution of this problem, we point out two possibilities.

\vspace{.3cm}
One possible way is to consider the interaction between the bulk and the probe D7 branes,
in which the electric field is imposed. This is equivalent to include $1/N_c$
corrections to the theory. We expect that this correction deforms the bulk
solution to the type of AdS-Schwartzschild solution which has a horizon. This
implies that the system get a low temperature as a result. Then, we can suppose
the configuration of the quark string connecting the D7 brane and the horizon.
But we do not perform this analysis here and it is remained as a future work.

The second idea of the resolution is as follows.
In the conductor phase of the present theory, the quark and the
anti-quark are living 
in the unstable ``mesons", which are called as the quasi-normal state
given in the next sub-section,
and
are supposed to be accelerated in the opposite direction. Such a string 
configuration is still expressed by the U-shaped one, and the both end points are
on the D7 brane. A time-dependent configuration, which would correspond
to this string-configuration, has been
found as an exact classical solution of the equation of motion of the fundamental
string in the AdS$_5$ space time \cite{Xiao}. 

This configuration is 
separated to three parts by dividing it at a special points of the radial
coordinate $(r=r_b)$ in AdS$_5$. The upper two parts for $r>r_b$ 
correspond to the quark and the anti-quark, which are moving with the velocity
below the speed of light. The speed of the lower part however exceeds the 
speed of the light. Then the causal parts of the quark and the anti-quark strings end 
at $r_b$. The observer on the boundary could see the disconnected
quark and anti-quark, which lose its energy into the lower part 
as the radiation of the accelerated charged particle \cite{Xiao}. 
This movement of the quark and anti-quark would be related to the current $B$.

This situation is very similar to the high temperature case, where
the causal parts of the quark and the anti-quark end at the horizon.
It would be possible to find a similar configuration also in the confining 
bulk background. }

While the configuration of this state is an U-shaped
string whose end points are on the 
probe brane, this can be changed to a static string with the Rindler temperature
by an appropriate coordinate transformation \cite{Xiao}. Then the accelerated configuration
mentioned above would be considered as an equivalent configuration given
at finite temperature within a coordinate transformation.


\vspace{.3cm}
{In the case of high temperature, deconfinement phase, we find the same
quasi-normal mode for the mesons in the conductor phase. Then it would be possible 
to see the same carrier of the current $B$ also in the deconfinement theory.
However, there is an event horizon in this case at $r_H$, so we could not see
the same carrier mentioned above when $r_H$ is larger than $r_b$. It would be an
interesting problem to study this point, but it is postponed as a future work
here.
}


\vspace{.9cm}
\noindent {\bf Numerical solutions:}

In the next, we show the numerical solutions in the conducting phase.
In solving numerically
the equation of motion(\ref{profile-eq}), we impose the boundary condition 
at the locus point as follows because of continuity of the solution $w(\rho)$,
\beq\label{bound}
\lim_{\epsilon\to 0}w(\rho_*-\epsilon)_{\rm in}=\lim_{\epsilon\to 0}w(\rho_*+\epsilon)_{\rm out},\ \ 
\ \ \lim_{\epsilon\to 0}w'(\rho_*-\epsilon)_{\rm in}=\lim_{\epsilon\to 0}w'(\rho_*+\epsilon)_{\rm out},
\eeq
where $w(\rho_*)_{\rm in(out)}$ represents the value slightly inside 
(outside) of the locus vanishing point. 

\begin{figure}[hbt]
 \begin{center}
  \includegraphics[width=90mm]{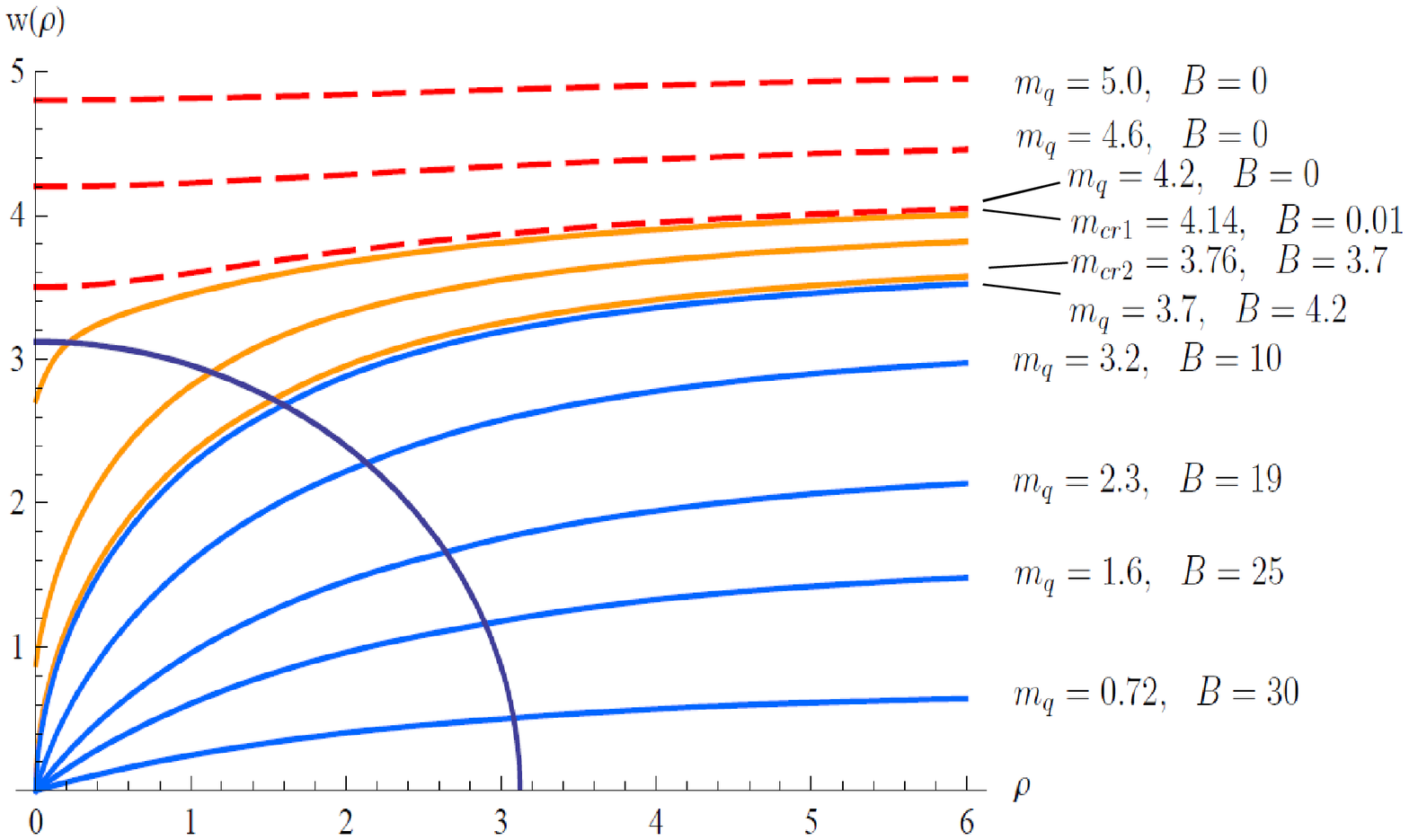}
\includegraphics[width=60mm,height=60mm]{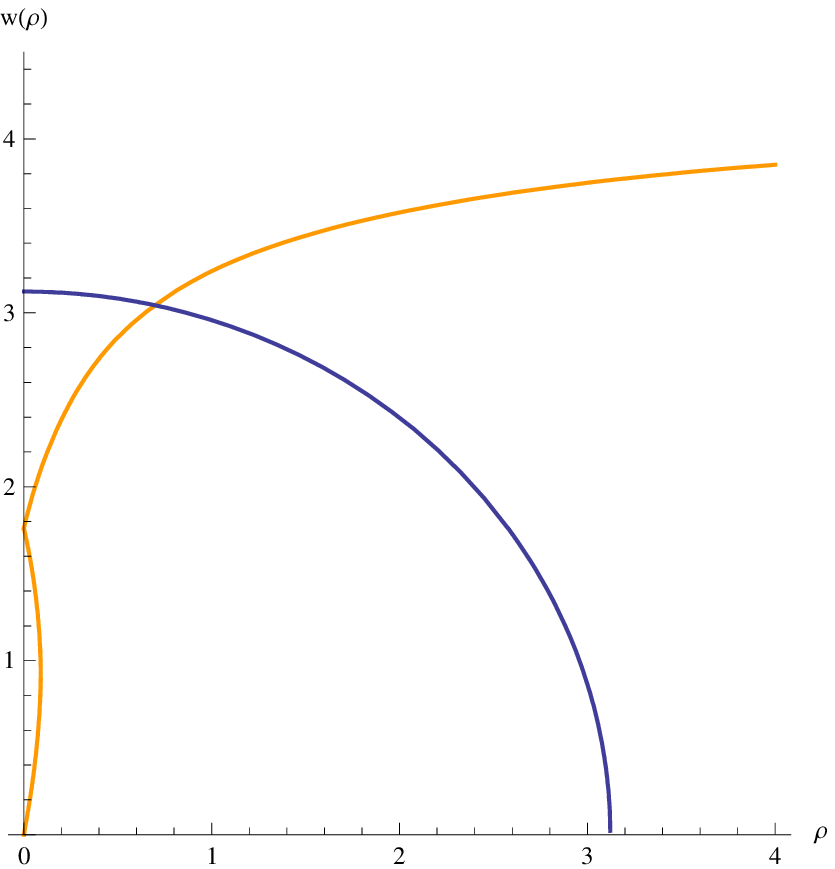}
 \end{center}
 \caption{{\small The solutions $w(\rho)$ for $\tilde{E}>q^{1/2}/R^2$. 
 Here we plot them for $\tilde{E}=10$, $q=5$ and $R=1$. 
Left: The upper dashed (red) curves represent the one of insulator phase 
 and they do not pass the locus vanishing point shown as a quarter of a circle. 
There are two types of solutions of the conductor phase. 
The middle (orange) ones end at finite $w(0)$. 
On the other hand, the lower (blue) ones end at $w(0)=0$. {Right: The actual conical singular solution of D7-brane. The conical
 singular solutions actually reflects at finite $w(0)$ and intersects
 to the event horizon $\rho=0$.}}}
 \label{fig:tb}
\end{figure}
%

The typical example of the numerical
results are shown in the Fig. \ref{fig:tb}. In general,
the solutions are separated to three groups. (i) For large quark mass, the solution
does not need $B$ since it does not cross the locus vanishing point. 
So the quarks are still in the insulator phase.

When the quark mass decreases, the second group solutions are obtained. 
The solutions $w(\rho)$ reach at the locus point, and these solutions 
demand finite $B$. In spite of the fact that the theory is in the confinement phase,
the carrier of this current can be considered as
the quarks and anti-quarks as explained above.

\vspace{.3cm}

\begin{figure}[hbt]
 \begin{center}
  \includegraphics[width=120mm]{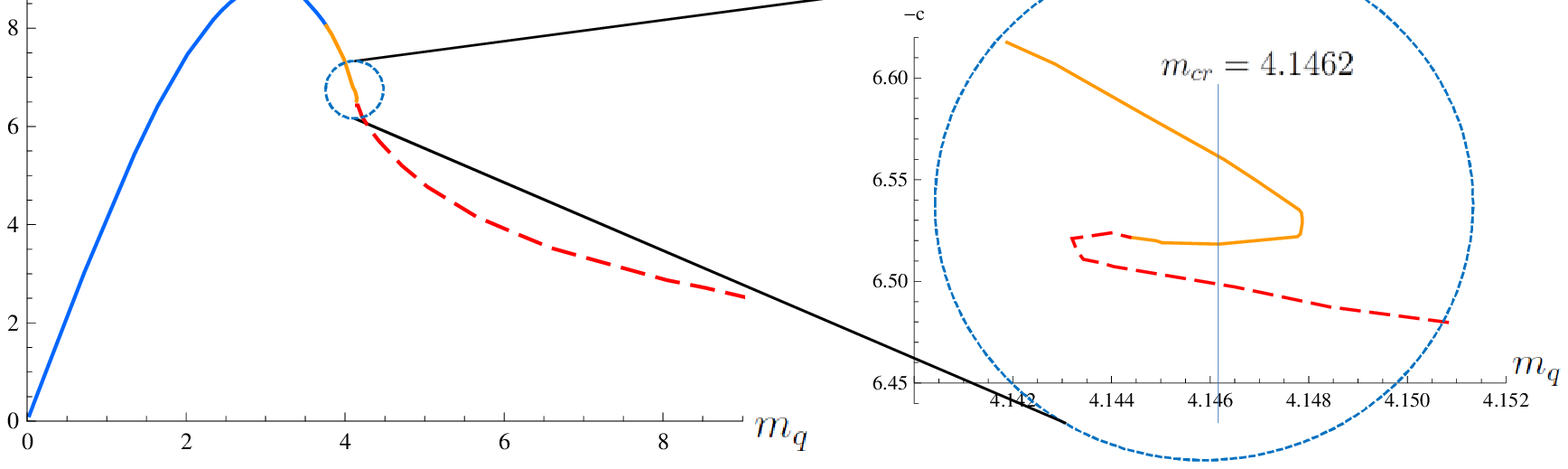}
 \end{center}
 \caption{{\small The relation between the chiral
condensation $c$ and the quark mass at $\tilde{E}=10$, $q=5$ and $R=1$. 
The solid curve includes two types of solution ((ii) and (iii)), 
namely blue curve ((iii)) and orange one ((ii)). 
The dashed (Red) curve represents the solution in the insulator phase. 
We can see a first order phase transition between insulator phase and conductor phase 
(from (i) to (ii)) from the right figure, where microscopic analysis is given.
The critical mass $m_{cr}$ for this phase transition is estimated by using the equal-area law.}}
 \label{fig:tc}
\end{figure}

Moreover, we find the solutions in the conducting phase
are separated to two types (see Fig.\ref{fig:tb}). 
{(ii) One satisfies $w(\rho=0)\neq 0$, and these solutions have a conical singularity. 
(iii) Another satisfies $w(\rho=0)\simeq 0$, which seems
smoothly reaching at the origin without any singular behaviour.}

\vspace{.3cm}
{The conical singular solution seems to be ending at finite $w(0)$
before reaching to the origin (here $r=0$), namely the event
horizon. The solution
which passes through $r^*$ must have the electric
current $B$ with its charge carriers, so the brane must end at the
horizon. Actually, we could find that the D7-brane of the 
category (ii) (classified as conical singular solution) 
could eventually reach to the horizon as shown in the right hand side 
of the Fig. \ref{fig:tb}. This result is consistent with the result 
obtained in the deconfining
case \cite{Albash:2007bk,Shock}. 

We don't have any reasonable physical interpretation for the solutions of
category (ii) from the gauge theory side. One possible resolution
for this singular solution is that we would need some stringy
corrections to remove this kind of solutions and to obtain more
smooth solutions, especially in the region of inside the locus
vanishing point $r<r^*$. For the region $r>r^*$, we consider however
that such corrections would be negligible and we could find physical insight.}

\vspace{.3cm}
In order to see the property of the transitions of these three types of solutions,
we studied the relation between quark mass $m_q$ and the VEV of quark bilinear $c=\langle \bar{\psi}\psi\rangle$, and
the numerical results are shown in the Fig. \ref{fig:tc}. 
For the transition from insulator to the conductor (from (i) to (ii)), the result shows
the multi-valuedness of $c$ as a function of 
$m_q$ near the transition point. 
This implies the typical behavior of the first-order phase transition.
On the other hand, for the transition from $w(0)\neq 0$ to $w(0)=0$ 
(from (ii) to (iii)), we could not find any
jump of $c$. 

\subsection{Quasi-normal mode and meson melting}
In the conductor phase, the solution $w(\rho)$ crosses the locus vanishing point.
In this case, we can see that there is no stable meson state since the energy
eigenvalue is complex. These are known as quasi-normal mode \cite{Starinets,Hoyos,evans}.
In the present confining theory, the same mode are found as shown below.

\vspace{.3cm}
{Here, we consider the meson spectrum in the conductor phase. 
For simplicity, we analyze the transverse components of the
vector mesons, $\delta A_{\bot}=\delta A_{y,z}$ 
given in the previous section (\ref{meson}), by imposing the following form of fluctuations,
\beq
\delta A_{\bot}=e^{-i\omega t}\xi(\rho).
\eeq
Then, the equation of motion of $\xi(\rho)$ is obtained:
\beq\label{meson-fluc}
{\cal G}^{\rho\rho}\xi''(\rho)+ 
\left(\partial_{\rho}{\cal G}^{\rho\rho}-2i\omega{\cal G}^{\rho t}\right)\xi'(\rho)+\left(-\omega^2{\cal G}^{tt}-i\omega\partial_{\rho}{\cal G}^{\rho t}\right)\xi(\rho)=0,
\eeq
where 
\bea
{\cal G}^{\rho\rho}&=&L_0{R^2\over r^2}\tilde{G}^{\rho\rho}\, \\
{\cal G}^{\rho t}&=&{\cal G}^{t \rho}=2L_0{R^2\over r^2}\tilde{G}^{\rho t}\, \\
{\cal G}^{tt}&=&L_0{R^2\over r^2}\tilde{G}^{tt}\,
\eea}
Here $B$ is retained as non-zero value,
and $\tilde{G}^{\rho\rho},\dots, \tilde{G}^{tt}$ are given in (\ref{inverse-metric}).

\vspace{.3cm}
According to \cite{Starinets}, one boundary is set at the locus
vanishing point, $r=r^*$ and $\rho=\rho^*$, 
where 
\beq
{r^*}^4e^{\Phi(r^*)}=R^4\tilde{E}^2\, , \quad {\rm and}\quad 
{\rho^*}^6e^{\Phi(r^*)}=R^2B^2\,
\eeq
are satisfied.
 
So the equation is firstly examined at this point.
Near this point the factors in the equation (\ref{meson-fluc})
are expanded by the powers of $\epsilon\equiv \rho-\rho^*(<<1)$ as follows,
\bea
{\cal G}^{\rho\rho}&=&a^{\rho\rho}_1\epsilon+O(\epsilon^2)\, \\
{\cal G}^{\rho t}&=&{\cal G}^{t \rho}=-a^{\rho t}_0+a^{\rho t}_1\epsilon+O(\epsilon)\, \\
{\cal G}^{tt}&=&a^{tt}_0+O(\epsilon)\, ,
\eea
where $(a^{\rho\rho}_1,\dots)$ are calculable finite coefficients.
The explicit form of these coefficients are abbreviated since they
are complicated. We like to notice one point that
$a^{\rho t}_0$ is positive. Then we find the leading part of the equation,
\beq\label{meson-fluc-2}
\xi''(\rho)+ {1\over \epsilon}\left(1+2i\omega{a^{\rho t}_0\over a^{\rho\rho}_1}\right)\xi'(\rho)+
{1\over \epsilon}\left(-\omega^2{a^{tt}_0\over a^{\rho\rho}_1}-i\omega
{a^{\rho t}_1\over a^{\rho\rho}_1}\right)\xi(\rho)=0,
\eeq
From equation (\ref{meson-fluc-2}), the solution near $\rho=\rho^*$
is obtained as the linear combination of two independent series,
\beq\label{solution}
 \xi=\alpha\epsilon^{-i\omega{a^{tt}_0\over a^{\rho\rho}_1}}\xi^{(1)}(\rho)
 +\beta \xi^{(2)}(\rho)
\eeq
where $\alpha$ and $\beta$ are arbitrary constants, and
$\xi^{(1)}(\rho)$ and $\xi^{(2)}(\rho)$ are expanded as
\beq
\xi^{(1,2)}(\rho)=a_0+a_1\epsilon+\cdots\, .
\eeq
The first term of the solution (\ref{solution}) represents the incoming wave
which should be
meaningful in the present case\cite{AFJK2}. 
So we consider the case of $\beta=0$.

\vspace{.3cm}
As shown in the high temperature model\cite{Starinets,Hoyos,evans},
this incoming
solution is connected to the local solution obtained at large $\rho$, where
we find also two independent solutions. 
In order to connect the solutions in the large and small $\rho$ regions,
the condition is taken 
such that the solution given above is smoothly connected to the normalizable
one. This condition provides discrete values of $\omega$, which are
complex, and the inverse of $\omega$ 
gives the decay constant of these modes. 
They are the quasi-normal modes mentioned above, and this implies that
the mesons melt into the quark and anti-quark in the conductor phase.

{In the present case, the role of the locus vanishing point seems to
plays as an effective horizon, though the bulk configuration has no
horizon and the locus vanishing point is made by the strong
electric field. Therefore the dynamical role of the electric field is
similar to the temperature. 
Then, we could propose the conical singularity does not effect
the phase diagram in the previous section because the
gauge theory could be ignorant of the physics inside of the locus
vanishing point.}



\section{Non-Supersymmetric case and Chiral Transition}
\begin{figure}[htbp]
\vspace{.3cm}
\begin{center}
 \includegraphics[width=7cm]{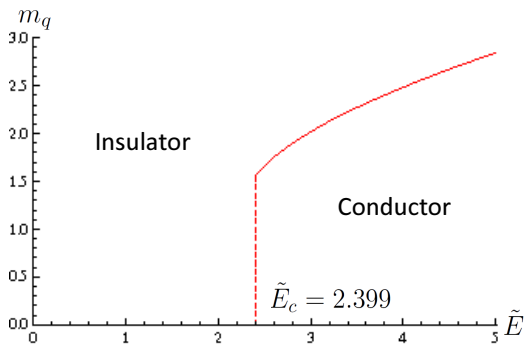}
\caption{{\small The phase diagram  for the non supersymmetric theory on the
 $m_q$-$\tilde{E}$ plane for  $r_0=1$ and $R=1$. The solid curve and dashed line represents the critical points
 between the insulator phase and the conductor phase. The dashed line is
 estimated by the minimum of $T(r)$ (\ref{insu-region}).}}
\label{mqetrfig}
\end{center}
\end{figure}

\vspace{.3cm}
Here we consider the solution given by (\ref{non-susy-sol}).
The electric field is added in a parallel way as given in the 
supersymmetric case.
The D7 brane is embedded in the world volume (\ref{D7-metric}), and the 
electric 
field is added by (\ref{electric-f}). Then,
after the Legendre transformation, the D7 brane Lagrangian with the 
electric field 
is given as
\begin{equation}\label{non-susy-action}
L_{D7}=\sqrt{(1+w'(\rho)^2)F_A(r)}\, ,
\end{equation}
where
\beq\label{non-susy-action2}
 F_A(r)={\rho^6}e^{2\Phi}A^8\left(1-{R^4\tilde{E}^2\over 
r^4A^4e^{\Phi}}\right)
\left(1-{R^2B^2\over \rho^6A^6e^{\Phi}}\right)\, .
\eeq

Then the equation of motion of $w$ is obtained from the above $L_{D7}$ as 
follows,
\beq\label{profile-eq2}
 \partial_{\rho}\left({w'\over \sqrt{1+{w'}^2}}\sqrt{F_A(r)}\right)
 -{w\over r}\sqrt{1+{w'}^2}{F_A'(r)\over 2\sqrt{F_A(r)}} 
 =0\, ,
\eeq
where $F_A'(r)=\partial_rF_A(r)$.

We solve this equation in two phases shown in the Fig. \ref{mqetrfig}.
The phase diagram is given in $m_q$-$E$ plane as in the case of the
supersymmetric case. It is similar to the one shown in the Fig. \ref{phase-diagram}
for the supersymmetric case, but we can not find two phases in the insulator side.
This point is explained below.


\subsection{Insulator phase}
\begin{figure}[htbp]
\vspace{.3cm}
\begin{center}
\includegraphics[width=7cm]{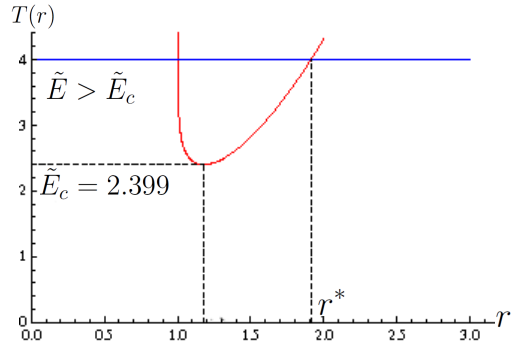}
\caption{{\small The solid (red) curve shows $T(r)\equiv A^2 e^{\Phi/2}r^2$ at
 $r_0=1$ and $R=1$. The minimum of $T(r)$ is shown by the dotted line at $T=2.399(=\tilde{E}_c)$. 
The locus vanishing point ($r^*$) for $\tilde{E}>\tilde{E}_c$ is
 obtained by the larger cross point between $T(r)$ (red curve) and the
 horizontal line at $T=\tilde{E}$ (solid blue line). 
\label{tensionfig}}}
\end{center}
\end{figure}

It is a little complicated to see the region of the insulator in the present
case. From Eqs.(\ref{non-susy-action}) and (\ref{non-susy-action2}), the insulator
phase is restricted to the region 
\beq\label{insu-region}
  \tilde{E}R^2\leq A^2e^{\Phi/2}r^2\equiv T(r)
\eeq
for any $r$. It is easy to see that $T(r)$ has a minimum in the region of
$r>r_0$. To understanding its behavior, we show $T(r)$ for $r_0=1$ and $R=1$
in the Fig. \ref{tensionfig}.

As mentioned, we find the minimum of $T(r)$, and we determine the
critical value of the electric field as $\tilde{E}\equiv \tilde{E}_c(=2.399)$
in terms of the inequality (\ref{insu-region}).
This inequality is satisfied for $\tilde{E}\leq \tilde{E}_c$
at any $r$. Then, the conductor phase appears in this region.

\begin{figure}[htbp]
\vspace{.3cm}
\begin{center}
\includegraphics[height=5cm,width=7cm]{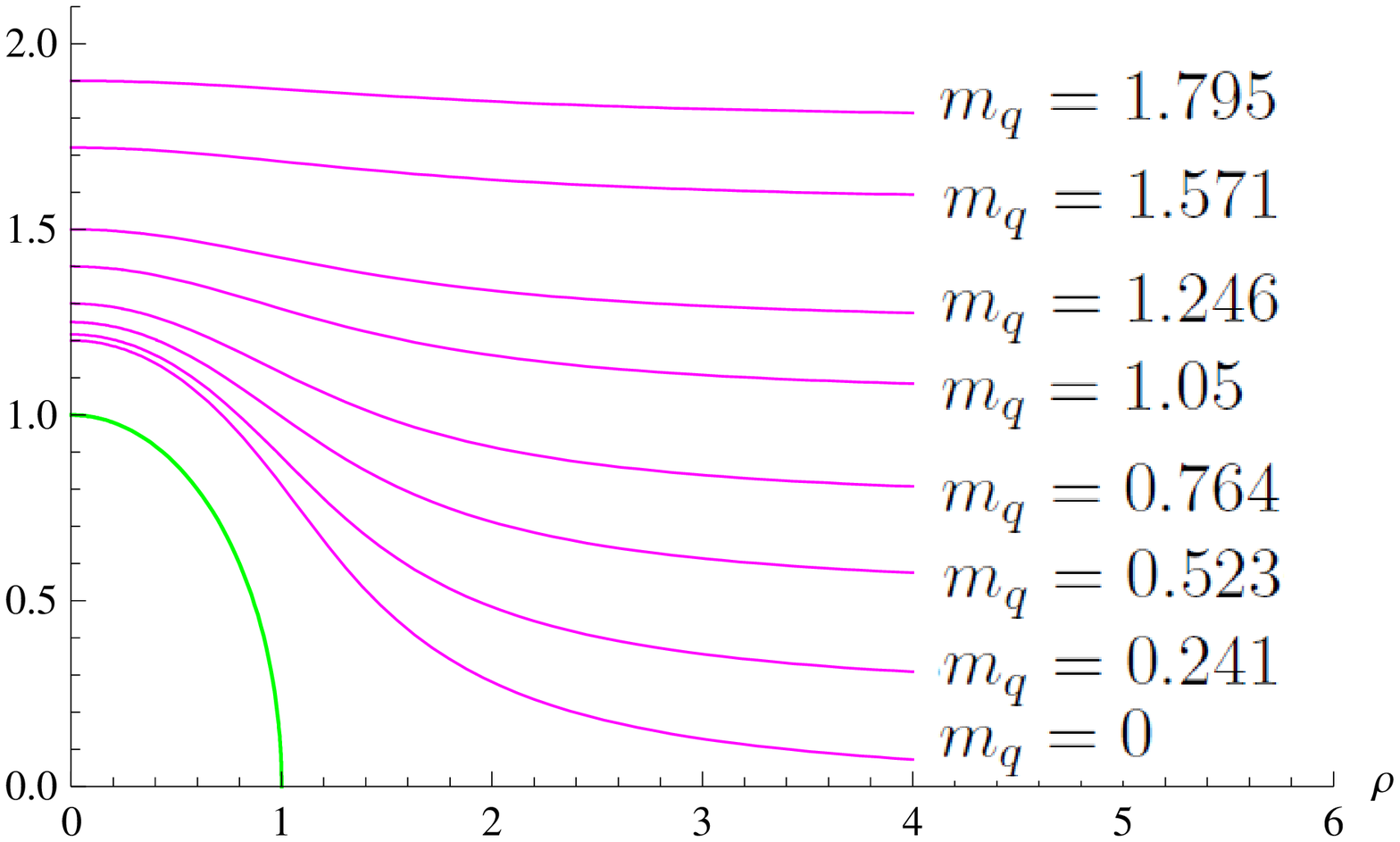}
\includegraphics[height=5cm,width=7cm]{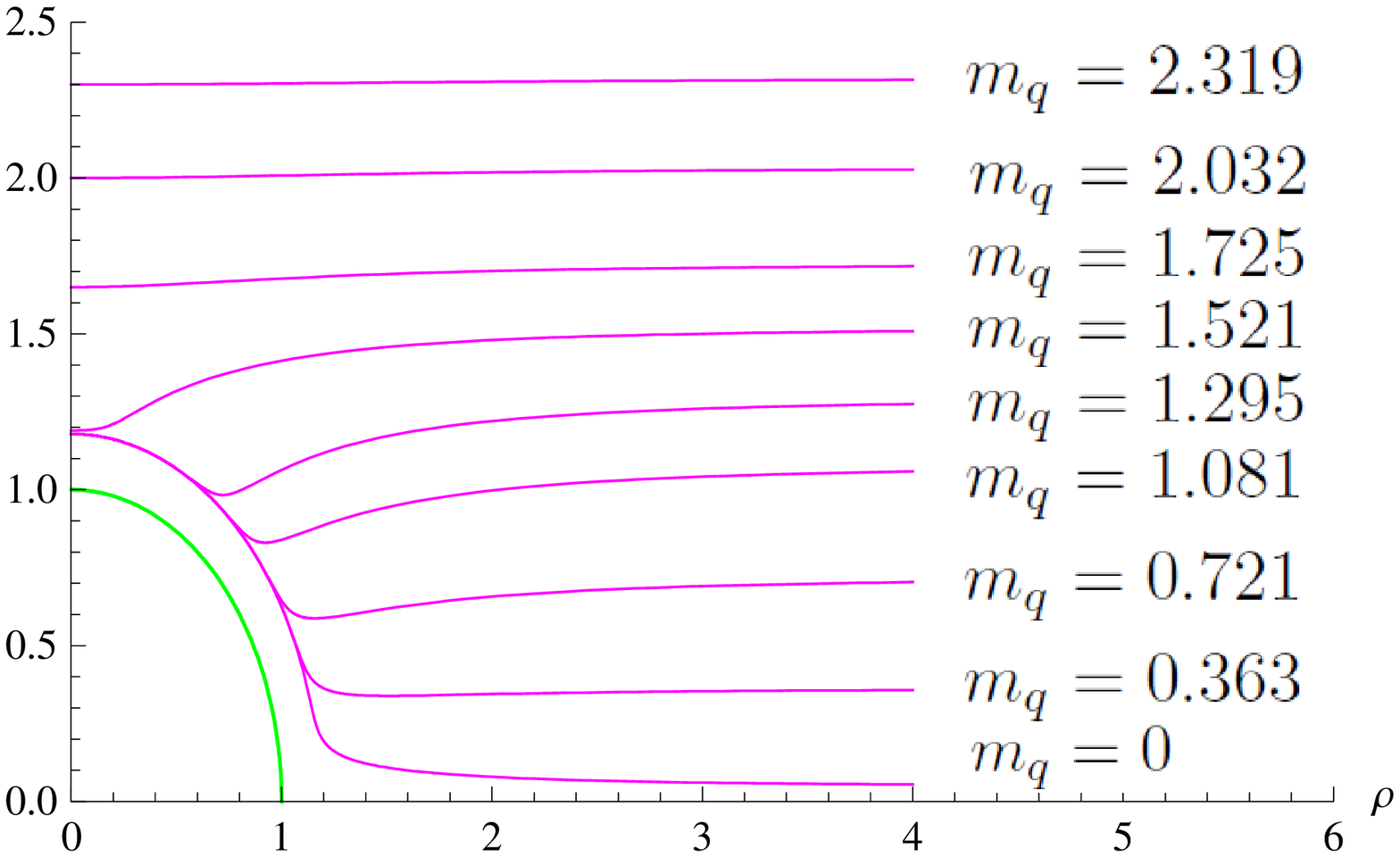}
\caption{{\small The solution $w(\rho)$ for various $m_q$ at $\tilde{E}=1.5$ (left
 figure) and $\tilde{E}=2.385$ (the right figure) $r_0=1$, and $R=1$.
The (green) circle represents the singularity at $r=r_0$.
}}
\label{ew15fig}
\end{center}
\end{figure}

\begin{figure}[htbp]
\vspace{.3cm}
\begin{center}
 \includegraphics[height=5cm,width=6.5cm]{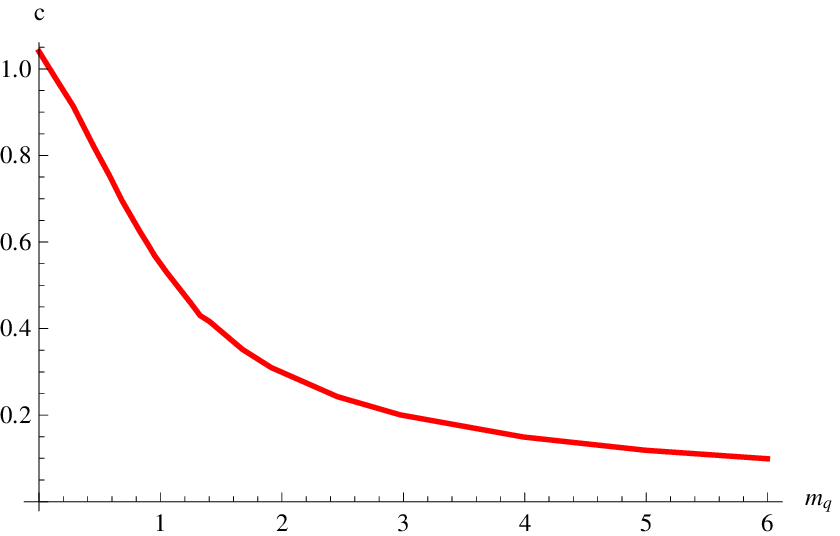}
 \includegraphics[height=5cm,width=6.5cm]{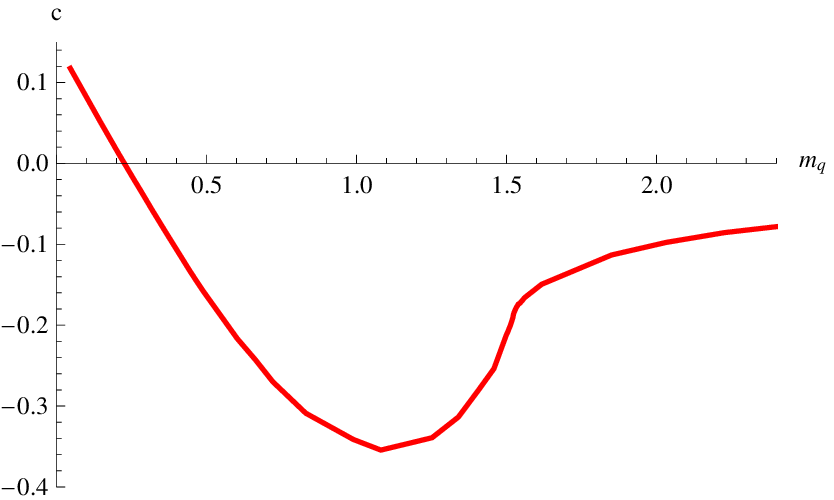}
\caption{The relation between quark mass $m_q$  and  the VEV of quark
 bilinear $c=\langle\bar{\psi}\psi\rangle$ for $\tilde{E}=1.5$ (the left
 figure) and $\tilde{E}=2.385$ (the right figure) at $r_0=1$ and
 $R=1$. }
\label{mqc15fig}
\end{center}
\end{figure}

We solve the equation of $w$ for the case for $\tilde{E}=1.5$ (far below $\tilde{E}_c$) and 
$\tilde{E}=2.385$ (just below $\tilde{E}_c$) in this insulator region. The solutions at these points 
give typical behavior of the insulator solutions.
The resultant solutions are shown in the
Fig. \ref {ew15fig} for various current quark masses,
$0\leq m_q\leq 1.795$ for $\tilde{E}=1.5$ (the left figure) and 
$0\leq m_q\leq 2.319$ for $\tilde{E}=2.385$ (the right figure)
respectively. For these parameters, the chiral condensate $c$s are shown for
each $m_q$ in the Fig. \ref{mqc15fig}.  

In the supersymmetric case, the insulator phase
has been separated to two phases (A) and (B) as shown in the Fig.\ref{phase-diagram}.
The transition between (A) and (B) is observed by the jump of the chiral condensate
$c$ at some $m_q$. 
However, as shown in the Fig. \ref{mqc15fig}, there is no such jump
in the present non-supersymmetric case. This is understood as follows. 
The end points $w(0)$ of
each solutions for various small quark-mass at $\tilde{E}=1.5$ do not degenerate as
shown in the case of the supersymmetric solutions, which are given 
in the Fig. \ref{ww}. This implies that there is no critical point of $m_q$,
where the value of the chiral condensate $c$ jumps.

\vspace{.3cm}
As $\tilde{E}$ approaches to $\tilde{E}_c$, the value of  $w(\rho=0)$ for small
quark masses closes to each other. For $\tilde{E}=2.385$, which is just below
of $\tilde{E}_c$, $w(\rho=0)$ for small quark mass seems to be almost 
degenerate as shown in the right of Fig.\ref{ew15fig}, 
but we could not find the jump of $c$ as shown in the Fig. \ref{mqc15fig}.

\subsection{Conductor phase}

Now we turn to the conductor solutions.
For $\tilde{E}\geq \tilde{E}_c$, there appears the locus vanishing point, $r=r^*$, 
which is determined for a fixed $\tilde{E}$ as 
\beq\label{locus}
 T(r^*)=\tilde{E}\, .
\eeq 
We could find two such points
from Fig \ref{tensionfig}, but here we define $r=r^*$ by the larger one.
As for the smaller cross-point given by (\ref{locus}), we do not consider
it since it exists very near to the singular point $r_0$, where the theory
requires many more higher curvature terms. We like to discuss
on this point in the future.
\begin{figure}[htbp]
\vspace{.3cm}
\begin{center}
\includegraphics[height=4cm,width=6.5cm]{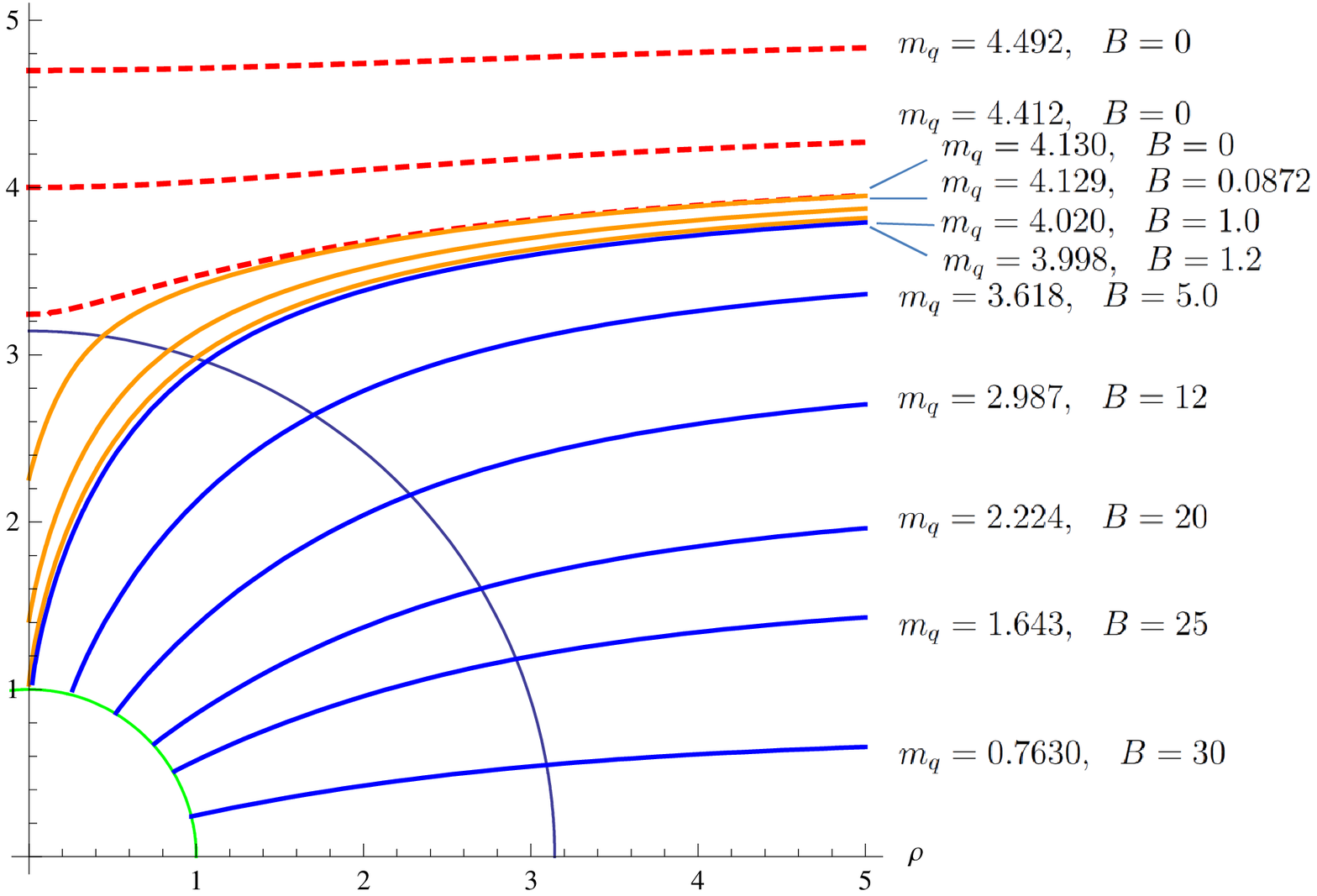}
\includegraphics[height=4cm,width=6.5cm]{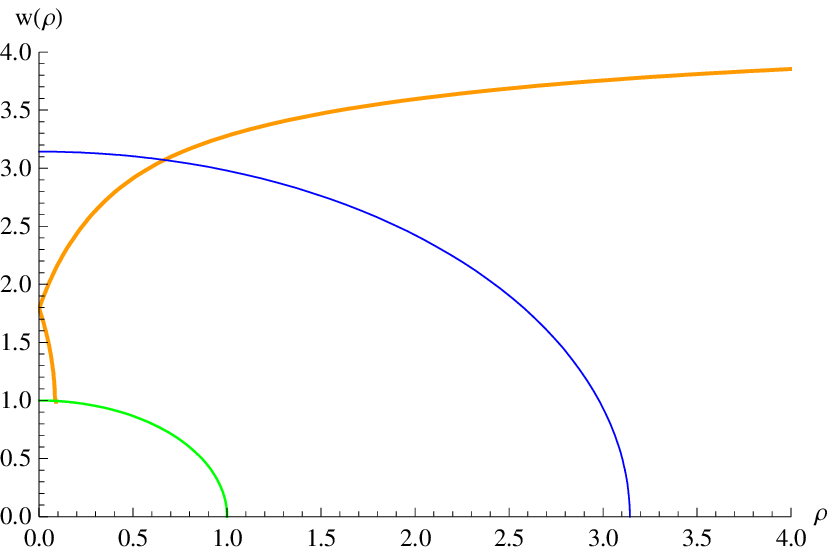}
\caption{{\small {Left figure:The solutions $w(\rho)$ are shown
for $\tilde{E}=10(>\tilde{E}_c)$, $r_0=1$ and $R=1$. 
The inner (green) and the outer (blue) circles
represent the singular point $r=r_0$  and the locus vanishing point $r=r^*$
respectively. Right: The conical singular solution of D7-brane is shown
 in detail.} }}
\label{e24fig}
\end{center}
\end{figure}
\begin{figure}[htbp]
\vspace{.3cm}
\begin{center}
 \includegraphics[height=4cm,width=6.5cm]{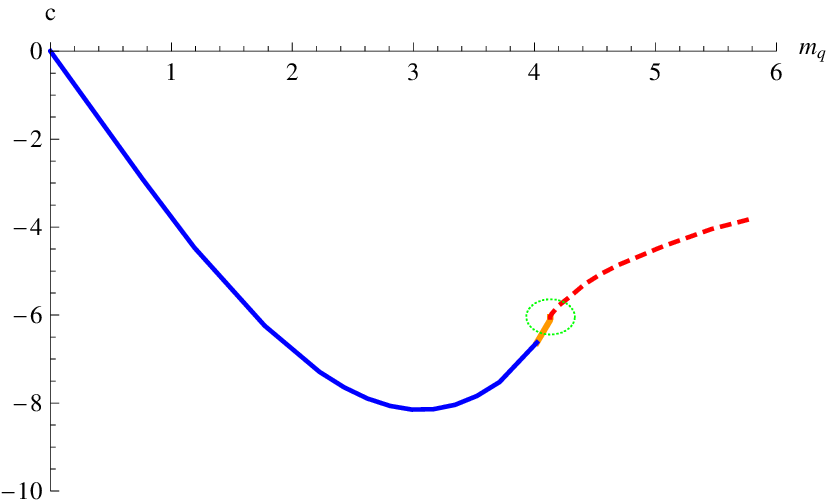}
 \includegraphics[height=4cm,width=6.5cm]{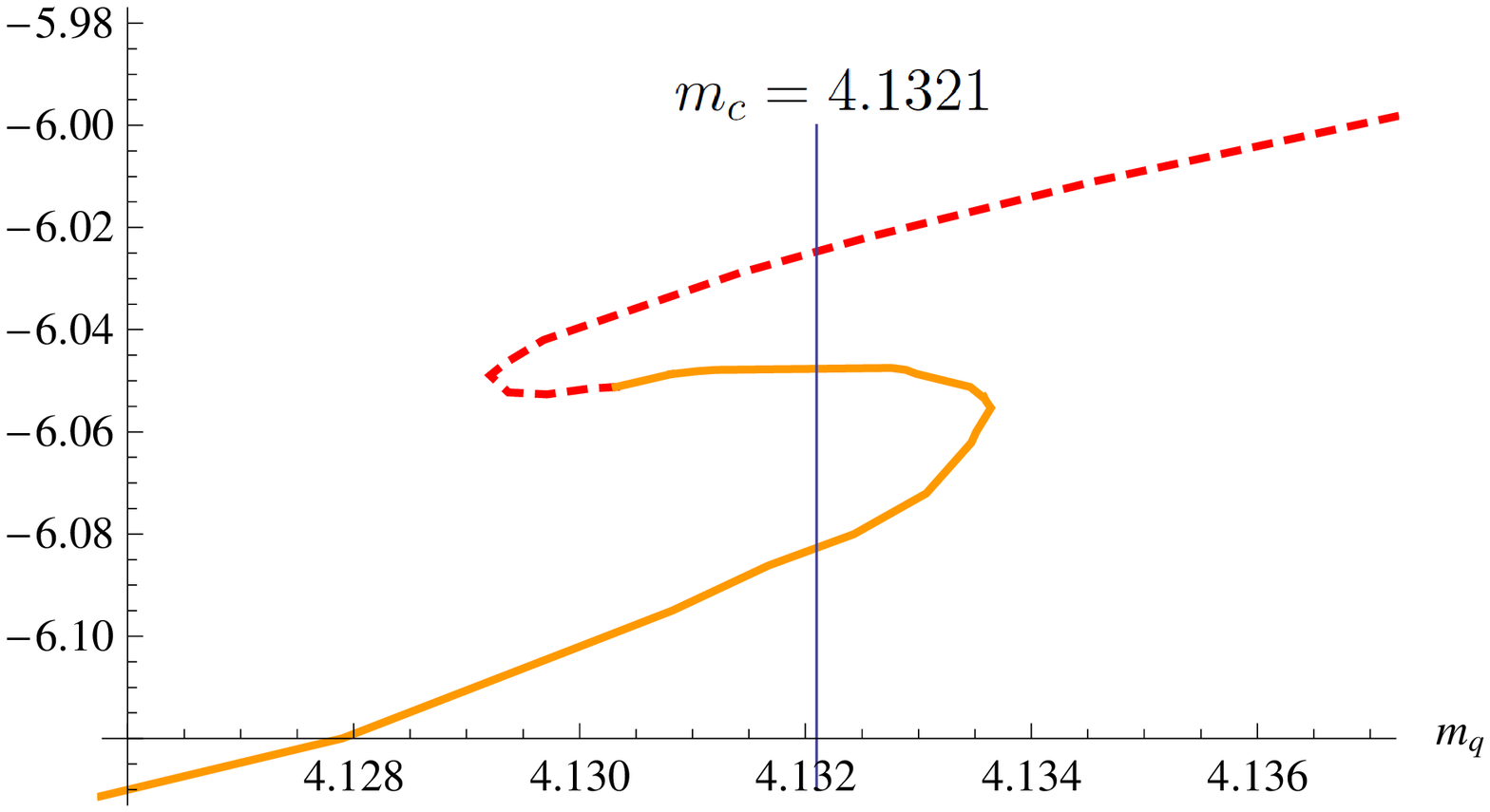}
\caption{{\small {The relation between condensation and the quark mass at
 $\tilde{E}=10$, $r_0=1$ and $R=1$. The solid (blue and orange) curves represent the solution in conductor
 phase. The orange curves represent the conical singular solutions. 
The dashed (red) curves represent the solution in the insulator phase. 
The critical mass $m_c=4.1321$ between conductor phase and insulator phase is
 estimated by using an equal-area law as shown in the right-hand figure.}}}
\label{e24mqcfig}
\end{center}
\end{figure}

For the conductor solutions,
$B$ is determined as
\begin{equation}
 B= {\rho^*}^3A^3(r^*)e^{\Phi(r^*)/2}/R\, ,
\end{equation}
then $w(\rho)$ could enter into the region $r<r^*$ in this case.
Thus
we solve the equation of motion for $w(\rho)$  by imposing the 
same boundary
condition (\ref{bound}) at the locus point as in the supersymmetric case.

\vspace{.3cm}
In order to study the typical solution, we study the case of
$\tilde{E}=10.0$, $r_0=1$ and $R=1$, and the results are shown in the Fig. \ref{e24fig}.
As expected, for large quark mass, we find
the solutions of the insulator phase. They are shown by the upper (red) curves, which 
are above the locus vanishing point $r^*$.
The middle (green) curves and the lower (blue) 
curves represent the solutions for the conductor phase. 
{The middle (orange) solutions and the lower (blue) one represent
the one for the conductor phase. 
The middle (orange) one are the conical singular solutions, and they
are bounced once on the $w$-axis then going down to the singular circle $(r_0)$
as shown in the right-hand figure of Fig \ref{e24mqcfig}. 
The lower (blue) solutions pass through the locus 
vanishing point and could approach smoothly 
to the point on the $r_0$.}

\vspace{.3cm}
In the next, we study the relation between quark mass $m_q$ and the VEV of
quark bilinear $c=\langle \bar{\psi}\psi\rangle$ near the transition
in order to see the property of the transition between different group
of solutions. In the Fig. \ref{e24mqcfig},
we show the results obtained near the transition
point between the insulator and the conductor. 
From this figure, we find the jump of $c$ at about $m_q\simeq 1.6$. 

\subsection{Chiral Transition}
Here we concentrate on the chiral condensate $c=\langle\bar{\psi}\psi\rangle$
for the massless quark. 
\begin{figure}[htbp]
\vspace{.3cm}
\begin{center}
\includegraphics[width=7cm]{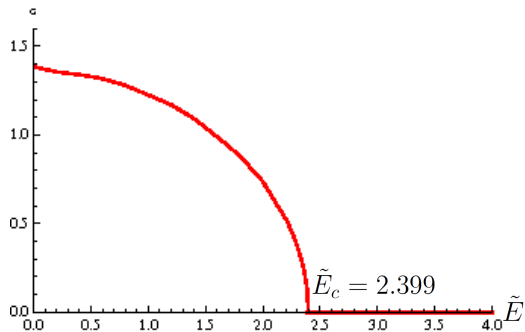}
\caption{{\small The relation between the VEV of chiral condensate 
$c=\bar{\langle\psi}\psi\rangle$ and $\tilde{E}$ for 
massless quark at $r_0=1$, $R=1$. $c$ becomes zero for $\tilde{E}\geq \tilde{E}_c=2.399$
}}
\label{ecfig}
\end{center}
\end{figure}
When $c=0$, we find chiral symmetry for the massless
quark, but it is spontaneously broken when we obtain $c\neq 0$.
In the present case, the chiral symmetry is broken at $E=0$.
The results of our calculation are shown in the Fig. \ref{ecfig}.
For small $E$, we have finite $c$, then
the chiral symmetry is spontaneously broken even if $E\neq 0$ in the
present non-supersymmetric confining theory \cite{GY}. However, $c$
vanishes in the conducting phase of $\tilde{E}>\tilde{E}_c$ as shown in the Fig. \ref{ecfig}.

Namely, the chiral symmetry is restored in the conducting phase where
the confinement force has been overcome by the strong electric repulsion.
In other words, the attractive force to generate 
finite VEV $c=\langle\bar{\psi}\psi\rangle$
has been eliminated by the electric field.
Similar result has been shown in \cite{Argyres}, where the analysis has been
performed in terms of a special form of string operator
based on the Sakai-Sugimoto model \cite{SSu} considered in the type IIA theory.
We notice here that the magnetic field has opposite role. 
Namely, it breaks the chiral
symmetry of the symmetric theory 
as shown by various holographic method 
\cite{Filev:2007qu,Zayakin,Kundu}.

{On the other hand, some analyses by the 4D field theories are also given
in \cite{Klev}. In that paper, the authors have studied the
Nambu-Jona-Lasino model with constant electromagnetic field and 
shown that chiral symmetry is restored above a certain critical
electric field strength. This result is  consistent with our results.
They have also shown the magnetic field breaks the chiral symmetry.}


\subsection{Meson melting and Quasi-normal mode}

As the above supersymmetric case, the quarks are confined also in the present
case, then we think the charge carriers as the quark and anti-quark
which are in a special state discussed in the supersymmetric case. 
Namely the carriers are formed of the melted mesons due to the strong electric field.

\vspace{.3cm}
{Here, we consider the meson spectrum in the conductor phase as in the supersymmetric
case. We analyze the vector mesons, 
$\delta A_{\bot}^{\rm NS}=\delta A_{y,z}^{\rm NS}$,  
where the notation "NS" is used to discriminate it from the supersymmetric case.
Then we express the fluctuations in the similar form to the supersymmetric case,
\beq
\delta A_{\bot}^{\rm NS}=e^{-i\omega t}\xi_{\rm NS}(\rho).
\eeq
And, the equation of motion of $\xi_{\rm NS}(\rho)$ is obtained:
\beq\label{meson-fluc-ns}
{\cal G}^{\rho\rho}_{\rm NS}\xi''_{\rm NS}(\rho)+ 
\left(\partial_{\rho}{\cal G}^{\rho\rho}_{\rm NS}-2i\omega{\cal G}^{\rho t}_{\rm NS}\right)\xi'_{\rm NS}(\rho)+\left(-\omega^2{\cal G}^{tt}_{\rm NS}-i\omega\partial_{\rho}{\cal G}^{\rho t}_{\rm NS}\right)\xi_{\rm NS}(\rho)=0,
\eeq
where 
\bea
{\cal G}^{\rho\rho}_{\rm NS}&=&L_0^{\rm NS}{R^2\over r^2A^2}\tilde{G}^{\rho\rho}_{\rm NS}\, \\
{\cal G}^{\rho t}_{\rm NS}&=&{\cal G}^{t \rho}_{\rm NS}=2L_0^{\rm NS}{R^2\over r^2A^2}\tilde{G}^{\rho t}_{\rm NS}\, \\
{\cal G}^{tt}_{\rm NS}&=&L_0^{\rm NS}{R^2\over r^2A^2}\tilde{G}^{tt}_{\rm NS}\,
\eea}
where
\bea
L_0^{\rm NS}&=&A^2{R\over r}\rho^3\sqrt{G_{(3)}^{\rm NS}}\, \\
\tilde{G}^{\rho\rho}_{\rm NS}&=&\left(A^4{r^4\over R^4}-\tilde{E}^2e^{-\Phi}\right)
/G_{(3)}^{\rm NS}\, \\
\tilde{G}^{\rho t}_{\rm NS}&=&-{{B}\tilde{E}R^2\over \rho^3 rA^6e^{\Phi}\sqrt{G_{(3)}^{\rm NS}}}\, \\
\tilde{G}^{tt}_{\rm NS}&=&-A^2\left(1+{w'}^2+{B}^2
  {G_{(3)}R^4\over A^{14}\rho^6 r^2e^{\Phi}}\right)/G_{(3)}^{\rm NS}\, \\
G_{(3)}^{\rm NS}&=&{r^2A^4\over R^2}(1+w'(\rho)^2){1-\frac{R^4\tilde{E}^2}{r^4A^4e^{\Phi}} 
\over {1-{{B}^2R^2\over \rho^6A^6 e^{-\Phi}} }}
\eea

These are different from the supersymmetric case by the factor $A(r)$ and the
form of the dilaton $e^{\Phi}$. In the present case, then, the locus vanishing point
$\rho^*$ is determined by (\ref{locus}). The point is slightly shifted
compared to the supersymmetric case. However, we should notice that $A(r)$ is not
singular at $r=r^*$. Then we could obtain the same qualitative behavior of the equation 
(\ref{meson-fluc-ns}) with the one obtained in the supersymmetric case
when we solve it near the locus vanishing point. 
This implies that there is an incoming waves which are expressed by the quasi-normal
modes considered above, then all the meson states in the conductor phase melt
into new quark and anti-quark state discussed above. Then we find carriers 
of the electric current.

\section{Summary}

The responses to the electric field are examined for two confining gauge theories
by the holographic approach. For enough strong $U(1)_B$ electric field 
($E>E_{\rm cr}$), the theories
are changed from insulator phase to the conductor phase. In one confining
theory, which preserves chiral symmetry and supersymmetry, we find two phases
in the insulator phase by studying the chiral condensate as the order parameter
which shows a jump at the transition point of these two phases. 
This transition is also assured by
the jump of the meson mass from a finite value
to a very small one for small quark masses.

However in the
other confining theory, in which chiral symmetry is spontaneously broken and
the supersymmetry is lost, the phase transition in the insulator phase,
which is observed in the supersymmetric theory, is not seen. 
Instead, we find chiral symmetry transition at the same point of insulator-conductor transition. 
For the massless quark, we find finite chiral condensate 
for zero and small electric field since the chiral symmetry is spontaneously broken
in this theory.
But it disappears when $E$ exceeds
the critical value ($E>E_{\rm cr}$),
then the chiral symmetry is restored at strong electric field.

\vspace{.3cm}
In both theories, as mentioned above, 
we find the insulator-conductor transition at $E_{\rm cr}$, which is obtained for
each theory. 
This transition is similar to the topological phase transition
from the Minkowski embedding to the Black hole embedding observed at in the finite
temperature theory. In this case, the two embeddings are discriminated by the 
infrared end points of the solutions whether it touches on the horizon or does not.
In the present case, the solutions of two phases are similarly
characterized by the 
behavior that the end {part goes through} the locus vanishing point or does not.
In any case, we could find the jump of the chiral condensate when the phase
changes from the insulator to the conductor 
as in the high temperature case. 

Thus we could find the conductor phase solutions, which demand a
finite electric current, for the strong electric field
($E>E_{\rm cr}$). However,
the theories examined here are in the quark confinement phase, then we should 
make clear the charge carriers in the conductor phase. The D7 configurations
in this phase go through the vanishing locus point. And the 
spectra of mesons show similar behavior to the one
obtained in the case of the high temperature D7-embedding solution
which ends on the event horizon. In this case, the frequencies of
the mesons are complex. These states
are known as the quasi-normal modes and they must decay to some melted state.
We could consider that they
melt into quark and anti-quark. However, in the present case,
we can not consider the same quark-string configuration with the one in the
high temperature phase, where the configuration of the quark is expressed
by a string which is connecting
the probe D-brane and the horizon. 
For the confining theory case, however, we can not consider this configuration
due to the lack of the event horizon.

\vspace{.3cm}
{Here we state two possibilities to resolve this point. 
One resolution is to consider the interaction between the D7 probe with 
electric field and the bulk background. This is corresponding to add $1/N$
corrections, we expect a deformation of the bulk to generate a kind of
event horizon or finite temperature as a result. In this case, we also be able
to find the carrier which could be seen at high temperature phase.

Another is to consider a new type of event horizon which exists on the string
when its end points on the D-brane are accelerated in the opposite direction.
For such a string configuration, the part below $r<r_b$ moves faster than 
the speed of light when we observe it from the boundary. This part
is called as ``radiation" part. Then the whole string configuration
can be separated into the quark and anti-quark and the radiation part
by the point $r_b$. And the energy
of the quark and anti-quark string parts flows into the radiation part.
We notice that, 
in our confining theory, the repulsion due to the electric
field becomes stronger than the attraction of the confining color force
in the conductor phase. 
This point implies that the mesons would melt into
the quark, anti-quark, and the radiation. Then we find the constant
current $B$ and its carriers.

}  
We need more new analysis to assure these picture, and 
we will give the details for these ideas in the 
near future.


\vspace{.3cm}
\section*{Acknowledgments}
The authors would like to thank F. Toyoda and A. Nakamura for useful discussions.
M. Ishihara thanks to N. Evans for useful discussions and
comments. The work of M. I. is supported by JSPS Grant-in-Aid Scientific
Research No. 20$\cdot$04335.



\newpage
\end{document}